\documentclass[useAMS,usenatbib]{mn2e}
\usepackage{epsfig}
\usepackage{color}
\usepackage{subfigure}

\def\Msun{\mbox{$M_\odot$}}

\def\lsim{\mathrel{\rlap{\lower3.5pt\hbox{\hskip0.5pt$\sim$}}
    \raise0.5pt\hbox{$<$}}}
\def\gsim{~\rlap{$>$}{\lower 1.0ex\hbox{$\sim$}}}

\def\Re{\mbox{$R_{\rm e}$}}

\def\mst{\mbox{$M_{\star}$}}
\def\SDM{\mbox{${\cal{S}}_{\rm DM}$}}
\def\St{\mbox{${\cal{S}}_{\rm tot}$}}
\def\gDM{\mbox{$g_{\rm DM}$}}

\newcommand{\goodgap}{\hspace{\subfigtopskip} \hspace{\subfigbottomskip}}

\title[SIM and DM scaling relations]{Secondary infall model and dark matter scaling relations in intermediate redshift early-type galaxies}
\author[V.F. Cardone et al.]{V.F. Cardone$^{1,2}$, A. Del Popolo$^{3}$, C. Tortora$^{4}$, N.R. Napolitano$^{5}$\\
$^1$I.N.A.F. - Osservatorio Astronomico di Roma, via Frascati 33, 00040\,-\,Monte Porzio Catone (Roma), Italy \\
$^2$Dipartimento di Scienze Fisiche, Universit\`{a} degli Studi di Napoli "Federico II",
Complesso Universitario \\ di Monte Sant'Angelo, Edificio N, via Cinthia, 80126 - Napoli, Italy \\
$^3$Dipartimento di Fisica e Astronomia, Universit\`{a} di Catania, Viale Andrea Doria 6, 95125 - Catania, Italy \\
$^4$Universit$\ddot{a}$t Z$\ddot{u}$rich, Institut f$\ddot{u}$r Theoretische Physik, Winterthurerstrasse 190, CH-8057, Z$\ddot{u}$rich, Switzerland\\
$^5$ INAF -- Osservatorio Astronomico di Capodimonte, Salita
Moiariello 16, I-80131 - Napoli, Italy}

\date{Accepted xxx, Received yyy, in original form zzz}

\begin{document}

\maketitle

\begin{abstract}

Scaling relations among dark matter (DM) and stellar quantities
are a valuable tool to constrain formation scenarios and the
evolution of galactic structures. However, most of the DM properties
are actually not directly measured, but derived through model
dependent mass mapping procedures. It is therefore crucial
to adopt theoretically and observationally well founded
models. We use here an updated version of the secondary infall
model (SIM) to predict the halo density profile, taking into
account the effects of angular momentum, dissipative friction and
baryons collapse. The resulting family of halo profiles depends on
one parameter only, the virial mass, and nicely fits the projected
mass and aperture velocity dispersion of a sample of intermediate
redshift lens galaxies. We derive DM related quantities (namely the
column density and the Newtonian acceleration) and investigate
their correlations with stellar mass, luminosity, effective
radius and virial mass.

\end{abstract}


\section{Introduction}

According to the concordance $\Lambda$CDM model \citep{CPT92},
dark energy (in the form of a cosmological constant or a varying
scalar field) and dark matter (hereafter, DM) are the dominant
actors on the cosmological scene \citep{WMAP5,P09,SNeIaSDSS}. In
particular DM represents most of
the total mass on galactic and cluster scales
and drives the formation and evolution of cosmic structures.
Roughly speaking, DM haloes form when the expanding
matter within (and surrounding) an overdense region experiences
deceleration because of the gravitational force, decouples from
the Hubble flow, collapses, and eventually, virializes. Models
including all these processes have been realized using different
approaches.

N\,-\,body simulations have been
the primary instrument to fully implement the nonlinearities
of the formation process which are realized in the dark halo
growth. Despite they do not allow to catch the full physics of the
galaxy formation, collisionless simulations have been successful in
reproducing a wide range of galaxy properties, e.g.
the spherically averaged halo density profile,
$\rho_{\rm DM}(r)$, which has been found to be
well described by a double power\,-\,law
relation with $\rho_{\rm DM} \propto r^{-3}$ in the outer regions and
$\rho_{\rm DM} \propto r^{-\alpha}$ at their centers with the exact
value of $\alpha$ remaining a matter of controversy. In the
popular NFW model \citep{NFW97} they find $\alpha = 1$ independently
on halo mass, while either a steeper $\alpha = 1.5$
(\citealt{Moore+98,Ghigna+00,FM01}) or even shallower values
(e.g., \citealt{Power+03, Fukushige+04, Navarro+04}) have been
claimed elsewhere. It is also possible that $\alpha$ is not
universal at all, but rather depending on halo mass, merger history
and substructures \citep{JS00,K01}.

On the contrary, semi\,-\,analytical models are
more flexible offering the possibility to include a vast variety
of physical ingredients. In particular, \cite{GG72}, \cite{Gott75}
and \cite{Gunn75} introduced the secondary infall model (SIM) to
describe the collapse and virialization of halos that are
spherically symmetric, have suffered no major mergers, and have
undergone quiescent accretion. After these first analysis, other
works have relaxed the assumption of purely radial
self\,-\,similar collapse by including non-radial motions arising
from secondary perturbations and taking care of both angular
momentum and stars to lead shallower or steeper density profiles
depending on the halo mass (see, e.g., \cite{DP10} and refs.
therein).

Numerical and semi\,-\,analytic
models generally lead to different predictions which need a
detailed observational scrutiny. Furthermore, the
adoption of the proper density model is the basic ingredient
to derive the global DM properties which are critical
parameters in the galaxy formation scenario: e.g., the
virial mass which is considered the driver of the heating process that
might affect the star formation history
(\citealt{Dekel_Birnboim06}; \citealt{Cattaneo08}) and the
overall star formation efficiency (\citealt{CW09}).

From this point of view, scaling relations among DM and stellar
quantities may provide an important test to constrain both the
formation scenarios and the DM properties. Early\,-\,type galaxies
(ETGs) are ideal tools for these aims. First of all they are
found to lie on the so-called fundamental plane
(FP) which tightly relates the central velocity dispersion, effective
radius and surface brightness. In particular, the well known deviation
(or ``tilt'') of the FP with respect to the expectation of the virial
theorem is still to be fully understood.
The different explanations proposed rely in turn on non homology,
variation of the stellar $M/L$
ratio with luminosity and varying DM content (see, e.g.,
\citealt{Busarello+97}, \citealt{Donofrio+06}, \citealt[T+09,
hereafter]{T09}). Each of these solutions may tell
a different story about the interplay between the DM and the stellar component,
thus, it is clear how the constraints on scaling relations can help
to shed light on formation and evolutionary processes.

Recently, there have been growing evidences that DM mass fraction
in the ETGs central regions is an increasing function of stellar
mass (or luminosity) hence supporting the idea that DM might be
the main driver of the FP
tilt (\citealt{Cappellari+06}, \citealt{Bolton+07},
\citealt{HB09}, \citealt{C09}, T+09, \citealt{Auger+10}). On the
other hand, the mean 3D DM central density has been found to decrease
with mass and luminosity (\citealt{Thomas+09}, T+09,
\citealt{Tortora+10}), while there are contradictory results on the
universality of the column density $\SDM = M_{\rm DM,proj}/\pi R^2$
(where $M_{\rm DM,proj}$ is the projected DM mass within the radius $R$)
with some results arguing for its constancy over 12 orders of
magnitude in luminosity (\citealt{D09,G09}) and other works
finding a correlation with halo mass \cite{B09}. Part of this
controversy may probably be ascribed to the different assumptions
on the halo model and stellar initial mass function (IMF) or the
adopted scale radius, as recently argued by some of us
\citep[hereafter CT10]{CT10}. On the other hand, it is also
possible that \SDM\ changes with the morphological type as
suggested by the recent results in \cite{NRT10}, where the central
projected density in ETGs is found to be, on average,
systematically higher than the same quantity for spiral and dwarf
galaxies (see also B09).

In order to further investigate this issue, we present here the analysis of
the above scaling relations based
on the SIM density profile obtained in Del Popolo (2010) adding to
the usual recipe of the gravitational collapse, the effects of ordered and random angular
momentum, dynamical friction and adiabatic contraction due to the baryonic collapse.

We use Einstein radius and velocity dispersion data from a
sample of intermediate redshift ($\langle z \rangle
\simeq 0.2$) lens galaxies from SLACS survey (\citealt[A+09,
hereafter]{Slacs}) to constrain the model parameters and derive
different scaling relations.
A general overview of the model is given in Sect. 2, while in Sect.
3 we introduce the lens sample and describe the fitting procedure.
Our main results are shown in Sect. 4 and discussed in the
concluding Sect. 5.

\section{The halo model}

The density profile of DM haloes is here obtained by
using the analytical method described in Del Popolo (2009,
hereafter DP09) which we refer the interested reader to for more
details. Here, we give a brief descriptions of the model
properties which are of main interest to our aims.

The halo profiles are derived by assuming the secondary infall
model (\citealt{GG72}) where a bound mass shell of initial
comoving radius $x_i$ expands up to a maximum radius (or
turnaround radius) $x_{\rm ta}$. As successive shells expand, they
acquire angular momentum and then contract on orbits determined by
the angular momentum itself, while dissipative processes and
eventual violent relaxation intervene to virialize the system
converting kinetic energy into random motions. The final density
profile may then be computed as\,:

\begin{equation}
\rho(x) = \frac{\rho_{\rm ta}(x_{\rm ta})}{(x/x_{\rm ta})^3} \left [ 1 + \frac{d\ln{(x/x_{\rm ta})}}{d\ln{x_{\rm ta}}} \right ]
\label{eq: defrhoend}
\end{equation}
with $\rho_{\rm ta}(x_{\rm ta})$ the density at turnaround and $x/x_{\rm ta}$
referred to as the collapse factor (see Eq. A18 in DP09). To
describe the proto--haloes density profile, DP09 considered the
profile of a peak in the density field generated according to the
Baarden et al. (1986) power spectrum and then took into account
angular momentum, dynamical friction and the presence of baryons
following the steps described below.

First, the angular momentum is decomposed in an ordered component,
related to the tidal torques experienced by proto--haloes, and a
random component connected to random velocities \citep{RG87}. The
ordered term is computed following Ryden (1988), while the random
part is assigned to proto--structures according to Avila\,-\,Reese
et al. (1998). A term related to the dynamical friction force has
been explicitly introduced in the equations of motion and
evaluated \cite{K80} dividing the gravitational force into an
average and a random component generated by the clumps in the
hierarchical universe. Finally, some adiabatic contraction of the
halo, due to the baryonic collapse, has been taken into account
through the formalism of Klypin et al. (2002) and Gnedin et al.
(2004), also including the exchange of angular momentum among
baryons and dark matter.

The final product of this halo formation method gives the DM
density profile as function of the radius $r$ and the total halo
mass $M_{\rm vir}$. The latter is the only parameter needed in
order to specify the halo density, being the halo inner slope
$\alpha$ a function of the virial mass as well. As discussed in DP10,
the dependence of $\alpha$ on $M_{\rm vir}$ breaks the universality
of the halo profiles and favour Burkert (1995) models at dwarf scales,
and models steeper than NFW at normal galaxy scales.

For the analysis we want to propose in the following, it is more
convenient to handle some analytical halo density profile, thus we
decided to approximate the numerical DP10 models with a generalized NFW
density profile \citep{JS00} which allows to accommodate varying inner slope
$\alpha$ being\footnote{The adoption of an analytical
approximation to the numerical output of the model will make the
adoption of the DP09 results easier to handle in the mass mapping
proposed below.}\,:

\begin{equation}
\rho_{\rm DM}(r, z) = \Delta_{\rm vir} \rho_{\rm crit}(z) \left ( \frac{r}{R_{\rm vir}} \right )^{-\gamma} \left ( 1 + \frac{c_{\rm vir} r}{R_{\rm vir}} \right )^{-(3 - \gamma)}
\label{eq: rhognfw}
\end{equation}
where $\rho_{\rm crit}(z) = 3H^2(z)/8 \pi G$ is the critical
density\footnote{We assume a concordance $\Lambda$CDM cosmological
model so that $H^2(z)/H_0^2 = \Omega_M (1 + z)^3 + (1 - \Omega_M)$
with $(\Omega_M, h) = (0.3, 0.7)$.} of the Universe at redshift
$z$ and $c_{\rm vir} = R_{\rm vir}/R_s$ is the halo concentration (with
$R_s$ and $R_{\rm vir}$ the radius where the logarithmic density slope
equals -2 and the virial radius respectively). Differently
from the original SIM we want to approximate (which is fully
assigned by the halo mass only), the generalized NFW profile is
formally a function of two parameters, namely $(c_{\rm vir},
M_{\rm vir})$. Thus, in Eq.(\ref{eq: rhognfw}), the dependence
on $c_{\rm vir}$ is fictitious and we can fit the same halo profile
with different $c_{\rm vir}$ values by changing the corresponding $\Delta_{\rm vir}$.
Due to this liberty in the $c_{\rm vir}$ choice, we therefore
arbitrarily scale $c_{\rm vir}$ with the total mass $M_{\rm vir}$ using
the popular relation \citep{B01}\,:

\begin{equation}
c_{\rm vir} = 12.81 \left ( \frac{M_{\rm vir}}{10^{12} \ {\rm M_{\odot}}} \right )^{-0.13} \ ,
\label{eq: cvmv}
\end{equation}
and then fit for $\Delta_{\rm vir}$ as function of $c_{\rm vir}$
obtaining\,:

\begin{equation}
\Delta_{\rm vir} \simeq \frac{6.626 \times 10^{-3}}{c_{\rm vir}^{0.1}}
\left \{ c_{\rm vir}^3 \left [ \ln{(1 + c_{\rm vir})} - \frac{c_{\rm vir}}{1 +
c_{\rm vir}} \right ] \right \}^{2.1} \ .
\label{eq: defdeltavir}
\end{equation}
With the chosen setup, the analytic expression of the inner
logarithmic slope $\gamma$ as a function of the mass $(M_{12} =
M_{\rm vir}/10^{12} \ {\rm M_{\odot}})$\, turns out to be written as:

\begin{equation}
\gamma \simeq 0.62 + \frac{1.166 M_{12}^{1/3} - 1}{1.166 M_{12}^{1/3} + 1} \ .
\label{eq: gammamv}
\end{equation}
We have checked that this analytical model fits extremely well the
numerical density profile over the mass range $10^{10} \le
M/M_{\odot} \le 10^{14}$ and is fully described by the virial mass as a single
parameter, as prescribed by the SIM numerical results.
Moreover, it is worth stressing that the use of the above
$c_{\rm vir}$\,-\,$M_{\rm vir}$ relation is just a convenient choice to
simplify the search for an analytical approximation which does not
affect the final accuracy of the numerical SIM profile fitting\footnote{In principle, one could have adopted a
whatever functional form provided the relation
$\Delta_{\rm vir}$\,-\,$c_{\rm vir}$ is adjusted in such a
way that the approximated density profile still fits the numerical
one. For this reason, one has not to update the
$c_{\rm vir}$\,-\,$M_{\rm vir}$ relation to account for a different
cosmology  or redshift. In particular, should one consider systems
with $z > 0$, one must still use  Eq.(\ref{eq: cvmv}) without
scaling $c_{\rm vir}$ by $(1+z)^{-1}$ as usually done in
literature.}.

As a final remark, we warn the reader that the above fitting
formulae for $\Delta_{\rm vir}$ vs $c_{\rm vir}$ and $\gamma$ vs $M_{\rm vir}$
have been obtained by considering haloes at $z = 0$ (since there
is a larger statistics and better resolution), while we will adopt
it also at the intermediate redshifts of the lenses we will
consider later. As can be seen from Fig. 2 in Del Popolo (2010),
the evolution of $\gamma$ with $z$ is actually quite small from $z
= 0$ to $z = 1$ over the mass range of interest here so that we
prefer to rely on these well checked approximations rather than
trying to fit less numerically accurate higher $z$ profiles.

\begin{table*}
\begin{center}
\scriptsize
\begin{tabular}{ccccccccc}
\hline
Lens id & $\log{L_V}$ & $\log{\Re}$ & $\log{\mst}$ & $\log{M_{\rm vir}}$ & $f_{\rm DM}(\Re)$ & $\log{\SDM(\Re)}$ & $\log{\gDM(\Re)}$ & $\log{g_{\star}(\Re)}$ \\
{}  & $\rm  \log L_{\odot}$ & $\rm kpc$ & $\rm \log{M_{\odot}}$ &
$\log{\rm M_{\odot}}$ &  & $\rm M_{\odot}/pc^2$ &
$\rm m/s^{2}$ & $\rm m/s^{2}$\\
\hline \hline SDSSJ0008$-$0004 & 11.11 & 1.01 & 11.64\, $\pm$ 0.14
& 12.96\, $\pm$ 0.63    & 0.46\,
    $\pm$ 0.16    & 3.16\, $\pm$ 1.10    & -9.67\, $\pm$ 0.20    & -9.63\,
    $\pm$ 0.14    \\ SDSSJ0029-0055 & 10.98 & 0.97 & 11.58\, $\pm$ 0.13    & 13.21\,
    $\pm$ 0.84    & 0.40\, $\pm$ 0.19    & 3.10\, $\pm$ 0.098    & -9.69\,
    $\pm$ 0.32    & -9.60\, $\pm$ 0.13    \\ SDSSJ0037-0942 & 11.16 & 0.94 & 11.73\,
    $\pm$ 0.06    & 13.67\, $\pm$ 0.30    & 0.37\, $\pm$ 0.18    & 3.17\,
    $\pm$ 0.10    & -9.53\, $\pm$ 0.28    & -9.38\, $\pm$ 0.06    \\ SDSSJ0157-0056 & 11.25 & 0.87 & 11.74\, $\pm$ 0.10    & 13.11\, $\pm$ 0.68
& 0.29\,
    $\pm$ 0.14    & 3.18\, $\pm$ 0.19    & -9.61\, $\pm$ 0.26    & -9.24\
    $\pm$ 0.10    \\ SDSSJ0216-0813 & 11.43 & 1.15 & 12.03\, $\pm$ 0.07    & 13.90\,
    $\pm$ 0.33    & 0.49\, $\pm$ 0.20    & 3.23\, $\pm$ 0.16    & -9.42\,
    $\pm$ 0.25    & -9.51\, $\pm$ 0.07    \\ SDSSJ0252+0039 & 10.85 & 0.76 & 11.46\,
    $\pm$ 0.13    & 11.96\, $\pm$ 1.1    & 0.17\, $\pm$ 0.09    & 3.05\,
    $\pm$ 2.21    & -10.00\, $\pm$ 0.21    & -9.30\, $\pm$ 0.13    \\ SDSSJ0330-0020 & 11.07 & 0.88 & 11.58\, $\pm$ 0.09    & 12.95\, $\pm$ 0.49
& 0.36\,
    $\pm$ 0.11    & 3.06\, $\pm$ 0.14    & -9.66\, $\pm$ 0.16    & -9.41\,
    $\pm$ 0.09    \\ SDSSJ0728+3835 & 11.04 & 0.83 & 11.69\, $\pm$ 0.12    & 12.19\,
    $\pm$ 0.96    & 0.17\, $\pm$ 0.08    & 2.92\, $\pm$ 0.36    & -9.92\,
    $\pm$ 0.19    & -9.21\, $\pm$ 0.12    \\ SDSSJ0819+4534 & 10.86 & 0.96 & 11.40\,
    $\pm$ 0.08    & 13.79\, $\pm$ 0.27    & 0.60\, $\pm$ 0.21    & 3.22\,
    $\pm$ 0.59    & -9.45\, $\pm$ 0.25    & -9.75\, $\pm$ 0.08    \\ SDSSJ0822+2652 & 11.09 & 0.97 & 11.69\, $\pm$ 0.13    & 13.55\, $\pm$ 0.64
& 0.40\,
    $\pm$ 0.21    & 3.16\, $\pm$ 0.27    & -9.57\, $\pm$ 0.33    & -9.48\,
    $\pm$ 0.13    \\ SDSSJ0903+4116 & 11.34 & 1.09 & 11.84\, $\pm$ 0.14    & 12.83\,
    $\pm$ 0.74    & 0.40\, $\pm$ 0.15    & 3.13\, $\pm$ 0.52    & -9.74\,
    $\pm$ 0.20    & -9.58\, $\pm$ 0.14    \\ SDSSJ0936+0913 & 11.04 & 0.90 & 11.68\,
    $\pm$ 0.12    & 12.93\, $\pm$ 0.76    & 0.29\, $\pm$ 0.13    & 3.21\,
    $\pm$ 3.04    & -9.73\, $\pm$ 0.22    & -9.36\, $\pm$ 0.12    \\ SDSSJ0946+1006 & 10.95 & 1.0 & 11.59\, $\pm$ 0.12    & 13.91\, $\pm$ 0.40
& 0.55\,
    $\pm$ 0.22    & 3.22\, $\pm$ 0.143    & -9.42\, $\pm$ 0.28    & -9.64\,
    $\pm$ 0.12    \\ SDSSJ0959+0410 & 10.44 & 0.53 & 11.15\, $\pm$ 0.06    & 13.43\,
    $\pm$ 0.37    & 0.27\, $\pm$ 0.13    & 3.16\, $\pm$ 0.13    & -9.53\,
    $\pm$ 0.27    & -9.15\, $\pm$ 0.06    \\ SDSSJ1016+3859 & 10.81 & 0.67 & 11.48\,
    $\pm$ 0.12    & 13.44\, $\pm$ 1.0    & 0.24\, $\pm$ 0.17    & 3.24\,
    $\pm$ 0.13    & -9.54\, $\pm$ 0.38    & -9.11\, $\pm$ 0.12    \\ SDSSJ1020+1122 & 11.13 & 0.81 & 11.80\, $\pm$ 0.12    & 13.12\, $\pm$ 0.77
& 0.20\,
    $\pm$ 0.12    & 3.14\, $\pm$ 0.65    & -9.66\, $\pm$ 0.29    & -9.06\,
    $\pm$ 0.12    \\ SDSSJ1023+4230 & 10.92 & 0.82 & 11.57\, $\pm$ 0.12    & 13.47\,
    $\pm$ 0.61    & 0.31\, $\pm$ 0.18    & 3.18\, $\pm$ 0.65    & -9.58\,
    $\pm$ 0.32    & -9.30\, $\pm$ 0.12    \\ SDSSJ1112+0826 & 11.12 & 0.88 & 11.73\,
    $\pm$ 0.08    & 14.00\, $\pm$ 0.34    & 0.43\, $\pm$ 0.20    & 3.31\,
    $\pm$ 0.14    & -9.31\, $\pm$ 0.26    & -9.27\, $\pm$ 0.08    \\ SDSSJ1134+6027 & 10.81 & 0.76 & 11.51\, $\pm$ 0.12    & 13.32\, $\pm$ 0.87
& 0.27\,
    $\pm$ 0.17    & 3.17\, $\pm$ 0.12    & -9.61\, $\pm$ 0.35    & -9.24\,
    $\pm$ 0.12    \\ SDSSJ1142+1001 & 10.96 & 0.89 & 11.55\, $\pm$ 0.08    & 13.37\,
    $\pm$ 0.46    & 0.36\, $\pm$ 0.17    & 3.13\, $\pm$ 0.12    & -9.62\,
    $\pm$ 0.28    & -9.44\, $\pm$ 0.08    \\ SDSSJ1153+4612 & 10.70 & 0.64 & 11.33\,
    $\pm$ 0.13    & 13.40\, $\pm$ 0.75    & 0.27\, $\pm$ 0.17    & 3.21\,
    $\pm$ 0.131    & -9.55\, $\pm$ 0.35    & -9.19\, $\pm$ 0.13    \\ SDSSJ1205+4910 & 11.1 & 0.96 & 11.72\, $\pm$ 0.06    & 13.76\, $\pm$ 0.28
& 0.43\,
    $\pm$ 0.19    & 3.21\, $\pm$ 0.10    & -9.46\, $\pm$ 0.26    & -9.43\,
    $\pm$ 0.06    \\ SDSSJ1218+0830 & 10.98 & 0.95 & 11.59\, $\pm$ 0.08    & 13.39\,
    $\pm$ 0.42    & 0.41\, $\pm$ 0.17    & 3.09\, $\pm$ 0.09    & -9.65\,
    $\pm$ 0.26    & -9.56\, $\pm$ 0.08    \\ SDSSJ1306+0600 & 10.82 & 0.84 & 11.43\,
    $\pm$ 0.08    & 13.92\, $\pm$ 0.28    & 0.54\, $\pm$ 0.19    & 3.28\,
    $\pm$ 0.12    & -9.32\, $\pm$ 0.22    & -9.48\, $\pm$ 0.08    \\ SDSSJ1313+4615 & 10.94 & 0.82 & 11.58\, $\pm$ 0.08    & 13.75\, $\pm$ 0.37
& 0.38\,
    $\pm$ 0.19    & 3.23\, $\pm$ 0.12    & -9.44\, $\pm$ 0.29    & -9.30\,
    $\pm$ 0.08    \\ SDSSJ1318-0313 & 11.14 & 1.20 & 11.67\, $\pm$ 0.090    & 13.47\,
    $\pm$ 0.32    & 0.59\, $\pm$ 0.19    & 3.12\, $\pm$ 0.29    & -9.70\,
    $\pm$ 0.26    & -9.96\, $\pm$ 0.09    \\ SDSSJ1402+6321 & 11.13 & 0.94 & 11.79\,
    $\pm$ 0.060    & 13.38\, $\pm$ 0.44    & 0.30\, $\pm$ 0.15    & 3.12\,
    $\pm$ 0.18    & -9.65\, $\pm$ 0.27    & -9.33\, $\pm$ 0.06    \\ SDSSJ1403+0006 & 10.82 & 0.77 & 11.44\, $\pm$ 0.08    & 13.10\, $\pm$ 0.83
& 0.29\,
    $\pm$ 0.14    & 3.11\, $\pm$ 0.11    & -9.67\, $\pm$ 0.29    & -9.34\,
    $\pm$ 0.08    \\ SDSSJ1416+5136 & 11.02 & 0.79 & 11.64\, $\pm$ 0.08    & 13.23\,
    $\pm$ 0.57    & 0.25\, $\pm$ 0.12    & 3.14\, $\pm$ 0.10    & -9.61\,
    $\pm$ 0.26    & -9.17\, $\pm$ 0.08    \\ SDSSJ1430+4105 & 11.27 & 1.07 & 11.93\,
    $\pm$ 0.11    & 13.86\, $\pm$ 0.50    & 0.44\, $\pm$ 0.21    & 3.22\,
    $\pm$ 0.30    & -9.44\, $\pm$ 0.30    & -9.44\, $\pm$ 0.11    \\ SDSSJ1436-0000 & 11.17 & 1.10 & 11.69\, $\pm$ 0.09    & 13.29\, $\pm$ 0.41
& 0.49\,
    $\pm$ 0.16    & 3.12\, $\pm$ 0.57    & -9.70\, $\pm$ 0.23    & -9.75\,
    $\pm$ 0.09    \\ SDSSJ1451-0239 & 10.84 & 0.77 & 11.39\, $\pm$ 0.06    & 13.44\,
    $\pm$ 0.37    & 0.35\, $\pm$ 0.15    & 3.12\, $\pm$ 0.11    & -9.59\,
    $\pm$ 0.26    & -9.39\, $\pm$ 0.06    \\ SDSSJ1525+3327 & 11.44 & 1.23 & 12.02\,
    $\pm$ 0.09    & 13.29\, $\pm$ 0.63    & 0.42\, $\pm$ 0.19    & 3.16\,
    $\pm$ 1.21    & -9.74\, $\pm$ 0.29    & -9.68\, $\pm$ 0.09    \\ SDSSJ1531-0105 & 11.12 & 0.96 & 11.68\, $\pm$ 0.09    & 13.84\, $\pm$ 0.36
& 0.47\,
    $\pm$ 0.20    & 3.22\, $\pm$ 0.25    & -9.44\, $\pm$ 0.27    & -9.48\,
    $\pm$ 0.09    \\ SDSSJ1538+5817 & 10.64 & 0.6 & 11.28\, $\pm$ 0.08    & 12.89\,
    $\pm$ 0.57    & 0.23\, $\pm$ 0.08    & 2.99\, $\pm$ 0.13    & -9.67\,
    $\pm$ 0.17    & -9.15\, $\pm$ 0.08    \\ SDSSJ1614+4522 & 10.82 & 0.94 & 11.47\,
    $\pm$ 0.12    & 12.66\, $\pm$ 0.70    & 0.42\, $\pm$ 0.13    & 3.06\,
    $\pm$ 1.82    & -9.79\, $\pm$ 0.16    & -9.65\, $\pm$ 0.12    \\ SDSSJ1621+3931 & 11.15 & 1.03 & 11.70\, $\pm$ 0.07    & 13.53\, $\pm$ 0.34
& 0.44\,
    $\pm$ 0.18    & 3.14\, $\pm$ 0.18    & -9.61\, $\pm$ 0.27    & -9.59\,
    $\pm$ 0.07    \\ SDSSJ1627-0053 & 11.01 & 0.84 & 11.70\, $\pm$ 0.09    & 13.63\,
    $\pm$ 0.51    & 0.31\, $\pm$ 0.17    & 3.20\, $\pm$ 0.13    & -9.50\,
    $\pm$ 0.31    & -9.23\, $\pm$ 0.09    \\ SDSSJ1630+4520 & 11.15 & 0.90 & 11.86\,
    $\pm$ 0.07    & 13.08\, $\pm$ 0.50    & 0.24\, $\pm$ 0.09    & 3.06\,
    $\pm$ 0.18    & -9.68\, $\pm$ 0.19    & -9.18\, $\pm$ 0.07    \\ SDSSJ1636+4707 & 10.99 & 0.83 & 11.63\, $\pm$ 0.08    & 12.69\, $\pm$ 0.6 &
0.25\,
    $\pm$ 0.08    & 2.98\, $\pm$ 0.41    & -9.74\, $\pm$ 0.16    & -9.26\,
    $\pm$ 0.08    \\ SDSSJ1644+2625 & 10.8 & 0.75 & 11.43\, $\pm$ 0.08    & 13.51\,
    $\pm$ 0.47    & 0.32\, $\pm$ 0.17    & 3.17\, $\pm$ 0.11    & -9.56\,
    $\pm$ 0.31    & -9.31\, $\pm$ 0.08    \\ SDSSJ2238-0754 & 10.836 & 0.766 & 11.45\,
    $\pm$ 0.06    & 12.94\, $\pm$ 0.85    & 0.27\, $\pm$ 0.12    & 3.05\,
    $\pm$ 0.11    & -9.71\, $\pm$ 0.24    & -9.32\, $\pm$ 0.06    \\ SDSSJ2300+0022 & 10.98 & 0.85 & 11.65\, $\pm$ 0.07    & 13.85\, $\pm$ 0.32
& 0.42\,
    $\pm$ 0.19    & 3.27\, $\pm$ 0.12    & -9.35\, $\pm$ 0.25    & -9.28\,
    $\pm$ 0.06    \\ SDSSJ2303+1422 & 11.11 & 0.98 & 11.71\, $\pm$ 0.06    & 13.66\,
    $\pm$ 0.31    & 0.42\, $\pm$ 0.19    & 3.15\, $\pm$ 0.10    & -9.55\,
    $\pm$ 0.27    & -9.49\, $\pm$ 0.06    \\ SDSSJ2321-0939 & 10.95 & 0.87 & 11.60\,
    $\pm$ 0.08    & 13.53\, $\pm$ 0.51    & 0.34\, $\pm$ 0.18    & 3.13\,
    $\pm$ 0.11    & -9.59\, $\pm$ 0.31    & -9.38\, $\pm$ 0.08    \\ SDSSJ2341+0000 & 11.06 & 1.15 & 11.73\, $\pm$ 0.08    & 13.25\, $\pm$ 0.41
& 0.50\,
    $\pm$ 0.15    & 3.89\, $\pm$ 11.2    & -9.74\, $\pm$ 0.21    & -9.80\,
    $\pm$ 0.08 \\
\hline SDSSJ0737+3216 & 11.344 & 1.20 & 11.96\,   $\pm$ 0.07    &
13.93\,   $\pm$ 0.28    & 0.58\,   $\pm$ 0.20    & 3.22\,
      $\pm$ 0.11    & -9.42\,   $\pm$ 0.22    & -9.67\,   $\pm$ 0.07    \\SDSSJ1100+5329 & 11.29 & 1.14 & 11.84\,
      $\pm$ 0.07    & 12.93\,   $\pm$ 0.46    & 0.44\,   $\pm$ 0.11    & 3.17\,   $\pm$ 1.52    & -9.76\,
      $\pm$ 0.15    & -9.67\,   $\pm$ 0.07    \\SDSSJ1106+5228 & 10.73 & 0.65 & 11.37\,   $\pm$ 0.06    & 13.90\,
      $\pm$ 0.30    & 0.39\,   $\pm$ 0.17    & 3.30\,   $\pm$ 0.142    & -9.30\,   $\pm$ 0.25    & -9.17\,
      $\pm$ 0.06    \\SDSSJ1204+0358 & 10.73 & 0.67 & 11.45\,   $\pm$ 0.060    & 13.67\,   $\pm$ 0.35    & 0.29\,
      $\pm$ 0.16    & 3.23\,   $\pm$ 0.13    & -9.45\,   $\pm$ 0.29    & -9.12\,   $\pm$ 0.06    \\SDSSJ1250+0523 & 11.16 & 0.85 & 11.77\,   $\pm$ 0.07    & 12.27\,   $\pm$
0.65    & 0.14\,   $\pm$ 0.06    & 3.13\,
      $\pm$ 2.99    & -9.95\,   $\pm$ 0.20    & -9.16\,   $\pm$ 0.07    \\SDSSJ2347-0005 & 11.33 & 1.0 & 11.83\,
      $\pm$ 0.08    & 14.00\,   $\pm$ 0.43    & 0.49\,   $\pm$ 0.22    & 3.32\,   $\pm$ 0.17    & -9.30\,
      $\pm$ 0.28    & -9.40\,   $\pm$ 0.08    \\
\hline

SDSSJ0044+0113 & 10.87 & 0.85 & 11.47\,   $\pm$ 0.09 & 14.22\,
$\pm$ 0.31     & 0.59\,   $\pm$ 0.21     & 3.38\,
      $\pm$ 0.15     & -9.20\,   $\pm$ 0.23     & -9.47\,   $\pm$ 0.09     \\ SDSSJ0912+0029 & 11.26 & 1.08 & 11.96\,
      $\pm$ 0.07     & 14.05\,   $\pm$ 0.32     & 0.50\,   $\pm$ 0.19     & 3.25\,   $\pm$ 0.13     & -9.36\,
      $\pm$ 0.23     & -9.44\,   $\pm$ 0.07     \\ SDSSJ0935-0003 & 11.52 & 1.31 & 11.96\,   $\pm$ 0.07     & 14.54\,
      $\pm$ 0.24     & 0.80\,   $\pm$ 0.14     & 3.39\,   $\pm$ 0.13     & -9.17\,   $\pm$ 0.16     & -9.89\,
      $\pm$ 0.07     \\ SDSSJ0956+5100 & 11.19 & 0.95 & 11.81\,   $\pm$ 0.08     & 14.09\,   $\pm$ 0.33     & 0.48\,
      $\pm$ 0.20     & 3.32\,   $\pm$ 0.152     & -9.27\,   $\pm$ 0.24     & -9.32\,   $\pm$ 0.08     \\ SDSSJ1143-0144 & 11.06 & 1.02 & 11.60\,   $\pm$ 0.09     & 14.15\,   $\pm$
0.28     & 0.65\,   $\pm$ 0.20     & 3.29\,
      $\pm$ 0.13     & -9.29\,   $\pm$ 0.21     & -9.68\,   $\pm$ 0.09     \\ SDSSJ1213+6708 & 10.92 & 0.87 & 11.49\,
      $\pm$ 0.09     & 14.20\,   $\pm$ 0.31     & 0.59\,   $\pm$ 0.21     & 3.36\,   $\pm$ 0.15     & -9.21\,
      $\pm$ 0.23     & -9.49\,   $\pm$ 0.09     \\ SDSSJ1719+2939 & 10.82 & 0.75 & 11.46\,   $\pm$ 0.08     & 14.05\,
      $\pm$ 0.32     & 0.47\,   $\pm$ 0.20     & 3.36\,   $\pm$ 0.15     & -9.23\,   $\pm$ 0.24     & -9.27\,
      $\pm$ 0.08     \\

\hline
\end{tabular}
\end{center}
\caption{Main stellar and DM quantities of the lens ETG sample adopted.
We first show lenses of the B sample, then the ones to be added to get the G and F samples.
Stellar parameters are
from the SLACS collaboration (A+09), while DM quantities are derived
by our best fit SIM model to the data. Columns are as follows\,:
1. galaxy ID, 2. V-band luminosity $L_{\rm V}$, 3. logarithm
of \Re, 4. logarithm of the total stellar mass as taken from A+09
(assuming a Salpeter IMF), 5. model virial mass, 6. 3D DM mass
fraction, 7. logarithm of the DM column density $\SDM(\Re)$, 8.
logarithm of DM acceleration, 9. logarithm of the Newtonian
stellar acceleration. All quantities are given with their
$1\sigma$ errors except $\log{L_V}$ and $\log \Re$ that have
negligible uncertainties.}
\label{tab: tabmass}
\end{table*}

\section{Testing the SIM model}\label{sec:SIMan}

There are two main characteristics of SIM halo model obtained
above that make it particularly interesting\,: {\it i.)} it
is theoretically well founded and intuitively incorporates most of
the dark and baryonic collapse physics, and {\it ii.)} it ends up
with a halo family which depends on a single parameter (the virial
mass) and can be written analytically as a generalized NFW (once
specified the dependence of the halo normalization on the NFW
parameter $c_{\rm vir}$).

As a first observational test for this halo model, we start
with a sample of ETGs for which we can use a
multi\,-\,technique approach as in CT10.
The sample includes 59
ETGs from the lenses catalog collected by the Sloan Lens ACS
(SLACS) survey (Auger et al. 2009, hereafter A+09) for which
the velocity dispersion $\sigma_{\rm ap}$ (within
a circular aperture of radius $R_{\rm ap}= 1.5 \arcsec$) and the
Einstein radius $R_E$, and hence the projected mass within it, $M_E
= M_{\rm proj}(R_E)$, are measured.
Following A+09, we will model the light distribution with a de
Vaucouleurs (1948)
profile with the effective radius \Re\ and total luminosity $L_V$
set to the values inferred from the $V$\,-\,band photometry.
Finally, we use the estimate of the total stellar mass, \mst, from
the SLACS team (A+09, Table 4) where a \cite{Salpeter55} IMF is assumed.

The median values of $R_{\rm ap}/\Re$ and $R_E/\Re$ (respectively 0.62
and 0.51) indicate that the data probe the
galaxy inner regions where we need to carefully account for the
stellar contribution to the model estimates of
$\sigma_{\rm ap}$ and $M_E$. To this end, we adopt the PS model
\citep{PS97} as its projection closely mimics the
Sersic (1968) surface brightness profile and provide an analytical
form of the physical quantities of interest.

Being the projected profile a Sersic model, the lensing properties
of the PS model are also analytically computed
\citep{C04,EM07}. In particular, the projected mass within $\xi =
R/\Re$ is\,:

\begin{equation}
\mst^{\rm proj}(\xi) = \mst \left [ 1 - \frac{\Gamma(b_n
\xi^{1/n})}{\Gamma(2n)} \right ] \label{eq: massprojsers}
\end{equation}
with \mst\ the total stellar mass, $\Gamma(x, y)$ and $\Gamma(x)$
the incomplete and complete $\Gamma$ functions and we set $n = 4$
to mimic the deprojected de Vaucouleurs (1948) profile used by the
SLACS team to fit the surface brightness of their galaxies.

The stellar and DM mass models are finally used as input for the
computation of the luminosity weighted
velocity dispersion profile.
As a first step, the line of sight velocity dispersion is given
by \citep{ML05}\,:

\begin{equation}
I(R) \sigma_{\rm los}^2(R) = \frac{G M_{\rm e}
\rho_{\star}^{e}}{\Upsilon_{\star}} \int_{\xi}^{\infty}
{\frac{K(\eta/\xi) \tilde{\rho}_{\star}(\eta)
\tilde{M}_{\rm tot}(\eta)}{\eta} d\eta} \label{eq: sigmalosgen}
\end{equation}
where $I(R)$ is the Sersic intensity profile, $\eta = r/\Re$, $M_{\rm e}$
and $\rho_{\star}^{e}$ are the total mass and the stellar density at
\Re, $\Upsilon_{\star}$ the stellar $M/L$ ratio, $M_{\rm tot}(\eta)$ is
the total mass, $K(\eta/\xi)$ a kernel function depending on the
choice of the anisotropy profile and the tilted
quantities are normalized with respect to their values at \Re.
We consider only isotropic models and take the
corresponding $K(\eta/\xi)$ from Appendix B of
Mamon \& Lokas (2005). The observed quantity is then obtained by
luminosity weighting $\sigma_{\rm los}$ in a circular aperture of
radius $R_{\rm ap}$. Note that, according to the SDSS survey strategy,
$R_{\rm ap}$ is fixed to $1.5 \ {\rm arcsec}$ so that $\xi_{\rm ap} = R_{\rm ap}/\Re$
changes from one lens to another.

Having set the stellar component quantities from photometry and
mass estimates, we are left with only one unknown parameter,
namely the halo mass $M_{\rm vir}$. We find its fiducial value
minimizing the merit function\,:

\begin{equation}
\chi^2 = \left [ \frac{\sigma_{\rm ap}^{obs} - \sigma_{\rm ap}^{th}(M_{\rm vir})}{\varepsilon_{\sigma}} \right ]^2
+ \left [ \frac{M_{\rm E}^{obs} - M_{\rm E}^{th}(M_{\rm vir})}{\varepsilon_{\rm E}} \right ]^2 \ ,
\label{eq: defchisq}
\end{equation}
where, as detailed in Cardone et al. (2009), the uncertainties
$(\varepsilon_{\sigma}, \varepsilon_{\rm E})$ are obtained by summing
in quadrature the observational errors and the theoretical
ones as derived from the propagation of uncertainties on
$(\mst, \Re)$. For each lens, the best fit value is
the one minimizing $\chi^2(M_{\rm vir})$, while $1$ (2) $\sigma$
confidence levels are obtained solving $\Delta \chi^2(M_{\rm vir}) =
\chi^2(M_{\rm vir}) - \chi^2_{\rm min} = 1.0$ ($4.0$).

In order to select
only the lenses with the higher significance of the model fit,
one can roughly set a threshold on the reduced $\chi^2$ values.
For our best-fit models, these are typically smaller than 1
(for a single degree of freedom -- two observational constraints
vs one model parameter) mainly because the inclusion of theoretical
errors on  $(\varepsilon_{\sigma},
\varepsilon_E)$\footnote{It is worth noting that such a noise term can
be hardly reduced. To understand why, let us consider the case of
the projected mass which is the sum of the stellar and DM terms.
The stellar term is proportional to the total stellar mass and
hence is known with an uncertainty obtained propagating those on
the luminosity and the stellar $M/L$ ratio. This later is actually
the major source of uncertainty in the final budget and cannot be
reduced unless one re-derive the estimate of $\Upsilon_{\star}$
relying on a larger set of colours than the one used by the SLACS
collaboration.}. We have therefore decided to be conservative and
define to split the sample on the basis of their observed
uncertainties rather than the significance of the fit and
define a {\it full} (F), {\it good} (G), {\it best} (B) as
follow.

The sample F contains all the 59 lenses. The G sample is made out
of 51 lenses with halo mass in the range $11 \le \log{M_{\rm vir}} \le
14$ and best fit values for $(\sigma_{\rm ap}, M_E)$ satisfying

\begin{displaymath}
-2.0 \le (\sigma_{\rm ap}^{obs} - \sigma_{\rm ap}^{bf})/\varepsilon_{\sigma} \le 2.0 \
\end{displaymath}

\begin{displaymath}
-2.0 \le (M_{\rm E}^{obs} - M_{\rm E}^{bf})/\varepsilon_{\rm E} \le 2.0 \ .
\end{displaymath}
The B sample is the most conservative one being made out of 46
lenses with mass in the same range, but following quality parameters

\begin{displaymath}
-1.0 \le (\sigma_{\rm ap}^{obs} - \sigma_{\rm ap}^{bf})/\varepsilon_{\sigma} \le 1.0 \
\end{displaymath}

\begin{displaymath}
-1.0 \le (M_{\rm E}^{obs} - M_{\rm E}^{bf})/\varepsilon_{\rm E} \le 1.0 \ .
\end{displaymath}

The overall agreement of the
best fit values of $(\sigma_{\rm ap}, M_E)$ with the observed ones is
always quite good with $rms(1 - \sigma_{\rm ap}^{bf}/\sigma_{\rm ap}^{obs})
\rangle \simeq 7\%$ $(5\%)$
and $rms(1 - M_{\rm E}^{bf}/M_{\rm E}^{obs}) \simeq 13\%$ $(9\%)$ for
sample F (B). As a consequence, most of the results we will
discuss in the following are quantitatively (within the errors)
independent on the sample adopted, thus we will refer to sample
G as the trade\,-\,off between fit quality and
improved statistics.

Looking more closely to the galaxy filling the F sample
(i.e. the bottom rows of Table \ref{tab: tabmass}), they mainly differ
from the G sample for their higher $\log{M_{\rm vir}}$, exceeding the
imposed limit of $10^{14} \Msun$ fulfilled by samples B and G.
This has been set as an upper limit for virial masses of typical
galaxy systems, and also to match the range of validity of our
analytical approximation of the SIM (see Sect. 2).

Such a large $\log{M_{\rm vir}}$ model estimates might be a warning
about the accuracy of the approximation adopted to convert the
SIM density profile in the one parameter generalized NFW which tend to
overestimate the virial masses. We have checked though that
all (but one) of the seven most massive systems filling the F sample
reside in overdense regions (following the definition adopted in Treu
et al. 2009) so that we cannot exclude that this mass excess in the
final virial mass estimate is real and given by the contribution
of the cluster itself which we can not constrain with our limited data
and is beyond the purpose of this analysis. Finally, these systems
turns out to have unreasonably large virial masses even if we would use
the NFW rather than the SIM model (see later).
Thus, we can confidently conclude that they might not represent a
critical issue specifically for the SIM model, but rather that the
lensing model is missing some external field to be tracked with
forthcoming more accurate analyses.
To be conservative, in the following, we will exclude these
very massive systems in all the scaling relation estimates.

Since we have only two observed quantities for each lens, we
expect to determine $M_{\rm vir}$ on a case\,-\,by\,-\,case basis with
some large uncertainty (see e.g.  Table \ref{tab: tabmass}). In fact,
according to the standard propagation of errors, most of the mass
related quantities (such as the DM mass fraction and the column
densities) will be known with poor precision. As a possible way
out, one could try to improve the constraints by binning the
galaxies according to luminosity or stellar mass and then fitting
for a parameter which is assumed to be the same for all the
objects in the same bin (e.g., \citealt{C09}). Unfortunately, this
is not possible here because of the non universality of the DM
mass profile. Indeed, since $M_{\rm vir}$ is obviously different from
one lens to another, the slope $\gamma$ and the concentration
$c_{\rm vir}$ will differ too so that the objects in the same bin
would have intrinsically different properties. As a consequence, we do not
attempt any binning and prefer to deal with large error bars
rather than introducing possible systematics in the analysis.

\subsection{SIM vs NFW and Burkert models}

The above analysis has demonstrated that the proposed SIM model is able to
fit the observed aperture velocity dispersion and projected mass within
the Einstein radius for the SLACS lenses sample with a reasonable significance.
One might want to check whether other ``standard'' halo density recipes can do
a comparable job. We have commented in Sect. 2 that our SIM model, because of
the dependence of the central slope on the virial mass, is able to match a
wide range of core behaviour from ``cuspy'' NFW to
``cored'' Burkert models. It is then interesting to compare the best-fit
results to $(\sigma_{\rm ap}, M_E)$ as performed in Sect. \ref{sec:SIMan}
and see whether data are able to favor one model with respect the other.

We need to first remark that the main difference among these three models
is that NFW and Burkert density profiles are intrinsically bi-parametric
(although there is a correlation between the two parameters in both cases,
see Sect. \ref{sec:monopar}) while SIM is mono-parametric by definition. According with
the Ockham's razor principle, this is an
important argument to take into account in our final considerations.
\subsubsection{SIM vs two--parameter NFW and Burkert profiles}\label{sec:bipar}
The two parameters of the NFW and Burkert profiles can be similarly
recast  in terms of the virial mass $M_{\rm vir}$ and the concentration
\footnote{Actually, the {\it concentration} definition strictly applies
to the NFW case. We decided to re-define it for the Burkert model
for convenience, although we do not assume that this has a defined
physical meaning.}
$c = R_{\rm vir}/R_{\rm s}$ where $R_s$
is the radius where the slope of the logarithmic density equals -2 in the
NFW case and the core radius in the Burkert one. Since we have only two
observed quantities for each lens, it is not possible to determine both
$(c, M_{\rm vir})$ on a case\,-\,by\,-\,case basis so that we must rely on
a different strategy. We use the results from CT10, where we have therefore
binned the
SLACS lenses in ten luminosity bins and expressed $(c, M_{\rm vir})$ in terms
of quantities that can be assumed to be equal for all lenses in the same
bin. The parameters  $(c, M_{\rm vir})$ are then determined a posteriori for
each single lens on the basis of the $\chi^2$ minimization adopted in Sect.
\ref{sec:SIMan}.

Both the NFW and Burkert models fit well the lens data, although,
as discussed in CT10, the Burkert model leads to quite small virial masses
and hence unexpectedly low virial $M/L$ values. In order to make a meaningful comparison, we use the same selection criteria defined above to define
{\it good} (G) and {\it best} (B) samples for the NFW and Burkert fits.
It turns out that the number of lenses in the G and B samples is $(52, 50, 56)$
and $(46, 25, 35)$ for the SIM, NFW and Burkert models, respectively.
Note that the same lenses enter the G samples for SIM and NFW with only
two cases excluded for the NFW model because of a larger virial mass.
This is a consequence of the fact that the SIM models results generally
fitted a range of masses (see Table \ref{tab: tabmass}) that, according to
the Eq.(\ref{eq: gammamv}), correspond to $\gamma\geq 1$, i.e. more
``cuspy'' systems.

On the contrary, almost all the lenses enter the G sample in the Burkert
case because they have a far lower virial mass and the three
missing lenses are excluded since they have $\log{M_{\rm vir}} \le 11$. When
one strengthens the constraint on the precision of the recovered
$(\sigma_{\rm ap}, M_E)$, the SIM model becomes clearly preferred, its B
sample being the richest one.

The SIM model turns out to be favored by the AIC and BIC statistics
\citep{L00} which measure the significance of the best--fit taking into account
the different number of degrees of freedom by penalizing the introduction
of unnecessary parameters. These two estimators are defined respectively as\,:

\begin{displaymath}
AIC = -2 \ln{{\cal{L}}_{\rm max}} + 2 n \ ,
\end{displaymath}

\begin{displaymath}
BIC = -2 \ln{{\cal{L}}_{\rm max}} + n \ln{N_{\rm data}} \ ,
\end{displaymath}
where ${\cal{L}}_{\rm max}$ is the maximum likelihood value, $n$ the number
of model parameters and $N_{\rm data}$ the number of constraints. Assuming
a Gaussian likelihood (as we have implicitly done), it is $-2
\ln{{\cal{L}}_{\rm max}} = \chi^2_{\rm min}$, while $n = 1 (2)$ for
the SIM (NFW and Burkert) model with $N_{\rm data} = 2$. The SIM
model turns out to have the lowest $AIC$ and $BIC$ values and then
the favoured with respect the NFW and Burkert ones, although the small
values of $\Delta AIC = AIC_{\rm SIM} - AIC_{\rm NFW}$ and $\Delta BIC =
BIC_{\rm SIM} - BIC_{\rm NFW}$ makes the NFW model statistically equivalent
in most cases (with the Burkert model generally excluded because of the
highest values).


A possible caveat is in order here. Although formally defined in the same
way, the $\chi^2$ values for the SIM, NFW and Burkert models, are not
fully equivalent since, for each lens, the total errors $(\epsilon_{\rm ap},
\epsilon_E)$ depend also on the adopted model because of the propagation
of the uncertainties on the stellar masses and effective radius (see CT10
for details\footnote{Summarizing, the difference between the theoretically predicted
and observed values of the velocity dispersion and projected
mass entering the $\chi^2$ evaluation are normalized with respect to the errors,
but these errors are partially model dependent. For instance, the uncertainty on the theoretically
predicted $M_{proj}$ is propagated from the one on the stellar mass through a formula which
depends on the adopted DM halo profile.}). Although such differences are actually quite small and do
not impact significantly the relative ranking of the models, we have
nevertheless computed the percentage best fit residuals and found (for
the good samples lenses)\,:


\begin{displaymath}
rms\left[\left ( \sigma_{\rm ap}^{obs} - \sigma_{\rm ap}^{th} \right )/\sigma_{\rm ap}^{obs} \right ] = \
7\% \ , \ 13\% \ , \ 14\% \ ;
\end{displaymath}

\begin{displaymath}
rms\left[\left ( M_{\rm E}^{obs} - M_{\rm E}^{th} \right )/M_{\rm E}^{obs} \right ] = \
12\% \ , \ 19\% \ , \ 16\% \ ;
\end{displaymath}
for SIM, NFW and Burkert models, respectively. Again, we find that
the SIM model best reproduces the observed data so that we can safely
argue that it should be preferred with respect to the NFW and Burkert
models. This is one of the central results of this paper, since we
seem to have proved that a mono-parametric density profile (including
most of the physics of the galaxy collapse) works better
than two-parameter density profiles with a much weaker physical
content (e.g. collapse and evolution of collisionless DM particles).
However, this result is based on a different procedure of fitting, thus
one can argue that the final significance of the fit might have been
affected by the use of binned data to constraint the halo parameters.
In the next Section we will use the well known correlations
between the two density parameters of NFW (and Burkert) profiles to
check whether their one--parameter re--writing can match the one--by--one
data with similar significance.

\subsubsection{SIM vs one--parameter NFW and Burkert profiles}\label{sec:monopar}
For the NFW density profile there is a well established correlation between
the virial mass and the concentration as found in N\,-\,body simulations.
Following the recent analysis in Mu$\tilde{{\rm n}}$oz\,-\,Cuartas et
al. (2011), we therefore set\,:

\begin{equation}
\log{c_{\rm NFW}} = a_{\rm NFW}(z) \log{\left ( \frac{M_{\rm vir}}{h^{-1} \ {\rm M_{\odot}}} \right )} + b_{\rm NFW}(z)
\label{eq: cvsmvirnfw}
\end{equation}
with

\begin{figure*}
\centering
\subfigure{\includegraphics[width=7.5cm]{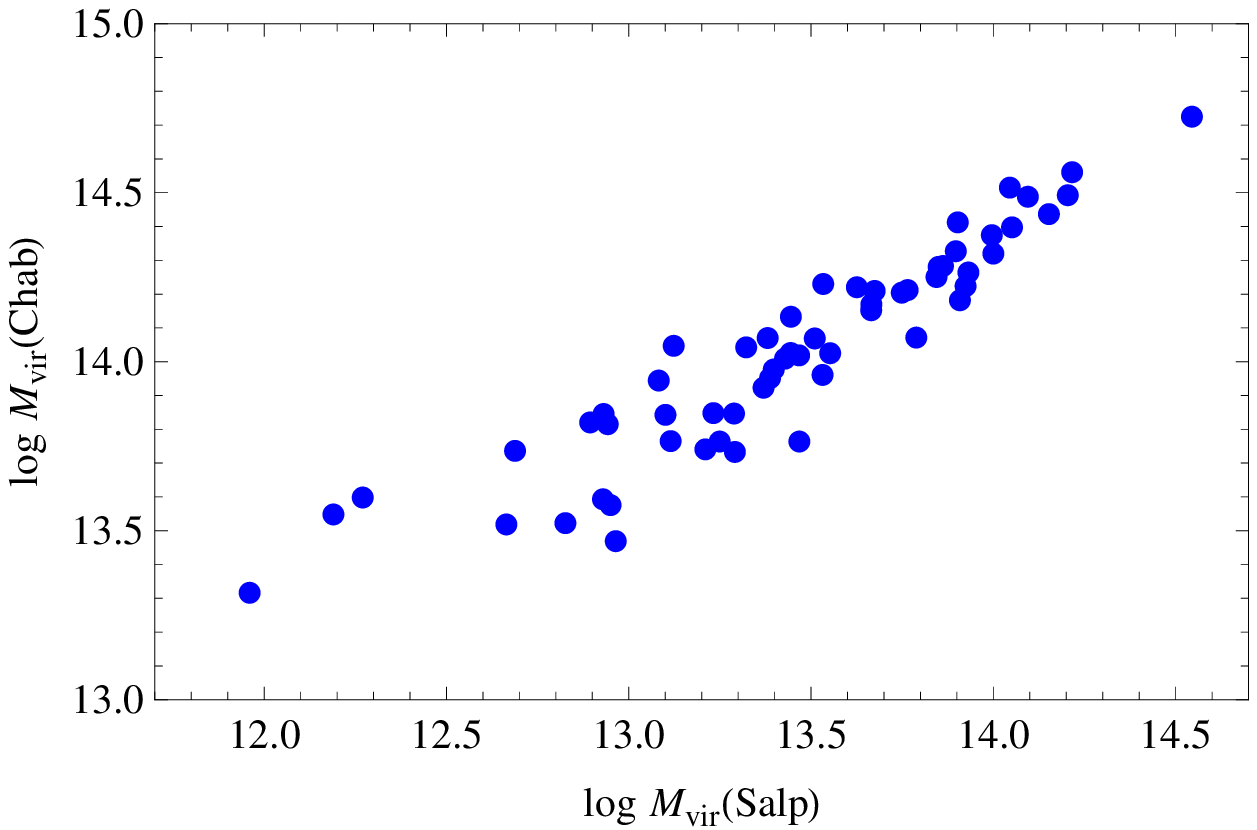}} \goodgap
\subfigure{\includegraphics[width=7.5cm]{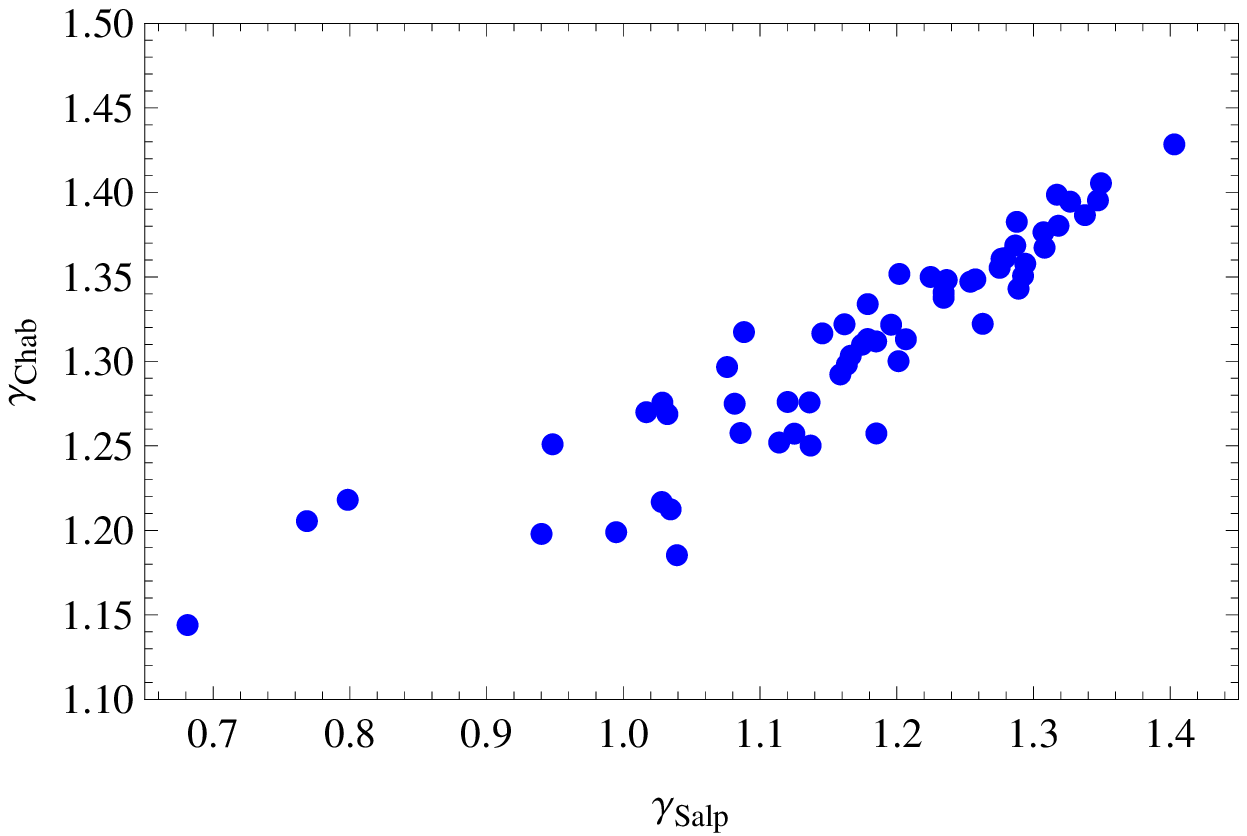}} \goodgap \\
\caption{Best fit virial mass (left) and inner slope (right) for the SIM model with Salpeter ($x$ axis) and Chabrier IMF ($y$ axis).}
\label{fig: salpchab}
\end{figure*}

\begin{displaymath}
a_{\rm NFW}(z) = 0.029 z - 0.097 \ ,
\end{displaymath}

\begin{displaymath}
b_{\rm NFW}(z) = - \frac{110.001}{z + 16.885} + \frac{2469.720}{(z + 16.885)^2} \ ,
\end{displaymath}
for $10 \le \log{[M_{\rm vir}/(h^{-1} \ {\rm M_{\odot}})]} \le 15$.
If we neglect the scatter in Eq.(\ref{eq: cvsmvirnfw}) as a first approximation,
this one parameter version of the NFW model can be fitted to the SLACS lenses
with the virial mass as only unknown quantity. Using this approach, we thus
find a viable solution for all the 59 objects in the sample with 53 (50)
lenses satisfying the selection criteria used to define the good (best)
sample for the SIM case. In particular, the quality of the fit may be quantified
by noting that

\begin{displaymath}
rms(1 - \sigma_{\rm ap}^{bf}/\sigma_{\rm ap}^{obs}) \simeq 4\% \ \  rms(1 - M_{\rm E}^{bf}/M_{\rm E}^{obs}) \simeq 13\%
\end{displaymath}
for the good sample and

\begin{displaymath}
rms(1 - \sigma_{\rm ap}^{bf}/\sigma_{\rm ap}^{obs}) \simeq 4\% \ \  rms(1 - M_{\rm E}^{bf}/M_{\rm E}^{obs}) \simeq 11\%
\end{displaymath}

for the best one.
We have repeated the same exercise with the Burkert model, by using
relation between the core density $\rho_0$ and the core radius $r_0$ found
in Salucci \& Burkert (2000) to reduce the density model to one--parameter
profile. Imposing this relation and fitting each single
lens using the virial mass as only parameter, we have found a viable
solution for only 6 lenses, which suggests that a mono--parametric
Burkert profile is uncapable to match the SLACS data and we do not discuss
this case hereafter\footnote{Note, however, that this is likely a
consequence of having assumed that
the $\rho_0$\,-\,$r_0$ relation found at $z = 0$ applies to any $z$.}.

If we go trough a more detailed comparison of the SIM to the
NFW model, we have seen that the latter works better in matching
the aperture velocity dispersion, while
the SIM is in a slightly better agreement with the projected mass values.
Overall, however, both models well match the data so that
the choice of which model is most viable should be driven by physical
motivations.
We indeed prefer the SIM model since the mass$-$central slope
relation comes out from a physical model rather than being the outcome
of numerical simulation. However, this argument alone does not allow us to
definitely abandon other density models and as far as both the SIM
and one parameter NFW (as well as the two--parameter version of NFW and Burkert
models, as discussed above) well reproduce the SLACS lens data, we will
hereafter consider all options and a posteriori check the difference introduced
by the different models.

\subsection{Changing from Salpeter to Chabrier IMF}

As fiducial values for the stellar mass of each lens, we have adopted the
estimates given in Table 4 of Auger et al. (2010) under the assumption of
a Salpeter IMF. However, since the IMF is still an uncertain variable in the
mass analysis, in this Section we want to investigating the impact of a different
IMF choice on the SIM results.

We have therefore consider the case of the Chabrier (2001) IMF
which returns stellar masses smaller by a factor 1.8 with respect
the ones obtained with the Salpeter IMF.
As the Chabrier IMF provides the lowest masses compatible with the
colors of the galaxy this will allow us to minimize the contribution
to the velocity dispersion and projected mass by the stellar component.

As for the Salpeter IMF, it turns out that all the lenses may be well
fitted with the best fit theoretical values of $\sigma_{\rm ap}$ and $M_E$
within $2 \sigma$ $(1 \sigma)$ of the observed ones for 59 (53) out of
59 lenses in the SLACS sample. Seemingly, the ability
of the SIM model to fit the data is not affected by the IMF choice.
However, when looking at the estimated virial masses for the
Chabrier IMF case, the inferred values
turned out to be generally larger (i.e., $\log{M_{\rm vir}} > 14.0)$
than the ones obtained by the Salpeter IMF such that only 25 (24) lenses
enter the G (B) sample.

This is made clearer in Fig.\,\ref{fig: salpchab}, where we show that
the best fit virial masses for the SIM\,+\,Chabrier model are almost
one order of magnitude  larger than those for the SIM\,+\,Salpeter case
because of the lower stellar masses predicted by the Chabrier IMF
which maximize the DM
contribution to the central parts. This is ``seen'' by the SIM model as
the presence of a more cuspy profile which implies larger $\gamma$ and
larger $M_{\rm vir}$ according to Eq.(\ref{eq: gammamv}).

Based on the argument of unrealistic virial masses we are inclined to
rule--out the SIM\,+\,Chabrier model. This is however becoming a common
conclusion from different analyses: e.g., gravitational lensing
(see, e.g., Treu et al. 2010, Cardone \& Tortora 2010) and studies of
the central DM fraction in local ETGs (Napolitano et al. 2011) suggest
that observations are consistent more with a Salpeter rather than a
Chabrier IMF for ETGs (unless some strong AC is considered).
Moreover, it has been also suggested that the IMF might vary with
luminosity \citep{RC93,T09} with a Salpeter one
being generally preferred for brighter systems (as probed by the
SLACS lenses).

\begin{figure*}
\centering \psfig{file=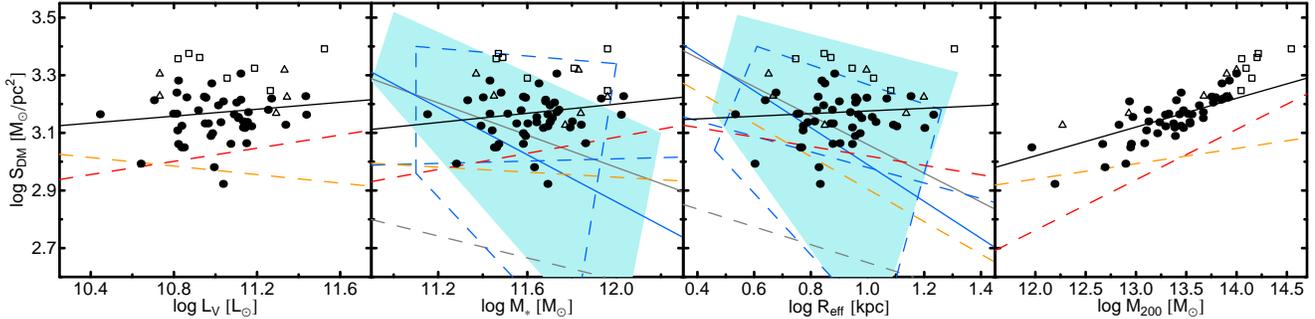, width=1\textwidth}
\caption{Best fit relations between the effective column density
$\SDM(\Re)$ and the total luminosity $L_V$, stellar mass \mst,
effective radius \Re\ and halo mass $M_{200}$ from left to right.
Data in sample B are plotted as points, while the other data not
present in B, but in samples G and F are plotted as triangles and
boxes, respectively. We do not plot error bars to not clutter the
plot. However, typical errors are $\langle
\sigma(\log{\mst})/\log{\mst} \rangle \simeq 1\%$, $\langle
\sigma(\log{M_{\rm vir}})/\log{M_{\rm vir}} \rangle \simeq 2\%$, $\langle
\sigma(\log{\SDM(\Re)})/\log{\SDM(\Re)} \rangle \simeq 11\%$,
while $\log{L_V}$ and $\log \Re$ have a negligible error. The
continue black line is the best fit made using sample G (i.e.
points and triangles). The red and orange dashed lines are the
best fit obtained in CT10, using a NFW + Salpeter IMF and a
Burkert + Salpeter IMF, respectively. The continue blue line and
cyan shaded region are the best fit and the region enclosing the
data for the results in \citet{Tortora+10} using a spherical
isothermal sphere, SIS (adopting a Salpeter IMF and projected
quantities), while the dashed ones are for the same data, but
using masses in A+09. Continue and dashed gray lines are the best
fits of local ETGs in T+09, using a SIS and a constant M/L
profile, respectively.} \label{fig: bfseff}
\end{figure*}

\section{Dark matter scaling relations}

Despite the large uncertainties, we have shown that the SIM halo model is able
to fit the combination of galaxy kinematics and lensing data fairly well.
We have also fixed the stellar mass contribution by assuming the Salpeter IMF,
we can now start looking into the DM properties of the galaxy sample.

The scaling relations we are interested in are commonly written as power
laws and can be
conveniently converted in linear relations in a log\,-\,log space,
$\log{y} = \log{A} + B \log{x}$ with comparable uncertainties on
$(\log{x}, \log{y})$.
As a best fitting procedure, we will follow the approach as in CT10
and adopt the Bayesian method described in D'Agostini (2005 see also Hogg et
al. 2010).

For the scaling relations we want to examine, we need to
choose a reference radius where the mass quantities are evaluated.
As these estimates will imply some model extrapolation, this choice
is critical sinc results can  be strongly model--dependent.

Since our data mainly probe the region close to \Re, this seems
the natural choice as the reference radius. Sometimes, also the
core radius $R_c$, introduced in the cored density models, and the
same $R_s$, where the logarithmic slope of the density profile is
$-2$, are taken as reference radius. It is, however, easy to check
that both these quantities are far larger than \Re\ so that one
would stay in the inconvenient position of deriving column density
estimates in regions much more far away of the distance
constrained by the data extension. As the proposed model has proven
itself to fairly well match the observations
around \Re\, we will compute the scaling relations at this reference distance.

\begin{table*}
\begin{center}
\begin{tabular}{|c|ccccc|ccccc|}
\hline
Fit & \multicolumn{5}{|c|}{Slope} & \multicolumn{5}{|c|}{Zeropoint}\\
\hline
~ & $x_{\rm BF}$ & $\langle x \rangle$ & $x_{\rm med}$ & $68\%$ CL & $95\%$ CL & $x_{\rm BF}$ & $\langle x \rangle$ & $x_{\rm med}$ & $68\%$ CL & $95\%$ CL \\
\hline \hline
$\SDM(\Re)$\,-\,$L_V$ & 0.06 & 0.02 & 0.02 & (-0.09, 0.14) & (-0.16, 0.23) & 3.1670 & 3.1671 & 3.1682 & (3.1631, 3.1694) & (3.1597, 3.1729) \\
$\SDM(\Re)$\,-\,$\mst$ & 0.08 & 0.06 & 0.07 & (-0.04, 0.17) & (-0.16, 0.28) & 3.1204 & 3.1228 & 3.1157 & (3.0740, 3.1782) & (2.9801, 3.2478) \\
$\SDM(\Re)$\,-\,$\Re$ & 0.05 & 0.07 & 0.07 & (-0.04, 0.19) & (-0.17, 0.30) & 3.1257 & 3.0788 & 3.0506 & (3.0117, 3.1882) & (2.9229, 3.2960) \\
$\SDM(\Re)$\,-\,$M_{200}$ & 0.10 & 0.10 & 0.10 & (0.05, 0.16) & (-0.02, 0.21) & 3.0189 & 2.9892 & 2.9615 & (2.9294, 3.0745) & (2.8660, 3.1560) \\
\hline \hline
$\St$\,-\,$L_V$ & -0.17 & -0.19 & -0.19 & (-0.27, -0.11) & (-0.37, -0.02) & 3.4390 & 3.4344 & 3.4364 & (3.4253, 3.4394) & (3.4245, 3.4424) \\
$\St$\,-\,$\mst$ & -0.16 & -0.16 & -0.16 & (-0.24, -0.07) & (-0.34, 0.02) & 3.5390 & 3.5207 & 3.5144 & (3.4777, 3.5747) & (3.4353, 3.6331) \\
$\St$\,-\,$\Re$ & -0.37 & -0.34 & -0.33 & (-0.45, -0.24) & (-0.59, -0.13) & 3.7635 & 3.7391 & 3.7487 & (3.6724, 3.8000) & (3.5613, 3.9340) \\
$\St$\,-\,$M_{200}$ & 0.01 & 0.01 & 0.01 & (-0.06, 0.08) & (-0.12, 0.16) & 3.4469 & 3.3381 & 3.3760 & (3.1088, 3.5102) & (3.1013, 3.6076) \\
\hline \hline
$\gDM$\,-\,$L_V$ & -0.04 & -0.02 & -0.01 & (-0.17, 0.12) & (-0.30, 0.28) & -9.628 & -9.625 & -9.626 & (-9.631, -9.615) & (-9.635, -9.614) \\
$\gDM$\,-\,$\mst$ & -0.03 & 0.00 & -0.01 & (-0.14, 0.13) & (-0.29, 0.29) & -9.607 & -9.621 & -9.611 & (-9.702, -9.550) & (-9.777, -9.441) \\
$\gDM$\,-\,$\Re$ & -0.04 & 0.03 & 0.03 & (-0.12, 0.18) & (-0.33, 0.35) & -9.590 & -9.637 & -9.625 & (-9.771, -9.535) & (-9.921, -9.332) \\
$\gDM$\,-\,$M_{200}$ & 0.25 & 0.24 & 0.24 & (0.17, 0.30) & (0.10, 0.37) & -9.947 & -9.978 & -9.961 & (-10.130, -9.864) & (-10.145, -9.770) \\
\hline \hline
$g_{\star}$\,-\,$L_V$ & -0.42 & -0.38 & -0.38 & (-0.51, -0.26) & (-0.63, -0.13) & -9.394 & -9.395 & -9.395 & (-9.397, -9.394) & (-9.399, -9.391) \\
$g_{\star}$\,-\,$\mst$ & -0.35 & -0.28 & -0.28 & (-0.41, -0.14) & (-0.58, 0.08) & -9.180 & -9.217 & -9.191 & (-9.287, -9.179) & (-9.422, -9.066) \\
$g_{\star}$\,-\,$\Re$ & -1.03 & -1.02 & -1.02 & (-1.12, -0.92) & (-1.24, -0.80) & -8.475 & -8.455 & -8.428 & (-8.529, -8.426) & (-8.647, -8.334) \\
$g_{\star}$\,-\,$M_{200}$ & -0.07 & -0.07 & -0.08 & (-0.15, 0.00) & (-0.25, 0.10) & -9.308 & -9.305 & -9.306 & (-9.378, -9.231) & (-9.527, -9.091) \\
\hline
\end{tabular}
\end{center}
\caption{Constraints on the slope and the zeropoint of the
correlations involving $\SDM(\Re)$, $\St$, \gDM\ and $g_{\star}$
using the sample G. For each parameter, we report the best fit,
mean and median values and the $68$ and $95\%$ confidence ranges.
Results from the fit to the other samples are fully consistent so
that we will not report them, but make them available on request.}
\label{tab: tabcorr}
\end{table*}

As a final remark, we stress that using $M_{\rm vir}$ as a model
parameter allows us to avoid extrapolation of the mass estimates
constrained at \Re\ out to the far larger virial radius, as
commonly done in parametric modeling procedures. This makes our virial
quantities intrinsically more robust.

\subsection{Column density}

We start by considering the correlations of the DM column density
$\SDM(R) = M_{\rm DM}^{\rm proj}(R)/\pi R^2$ with luminosity, stellar
mass, effective radius and virial mass. The best fit relations
obtained for the $\SDM(\Re)$ \footnote{Hereafter, we will discuss
only the results for the sample G having checked that fully
consistent constraints are obtained using the F and B samples.
This can also be qualitatively seen in the figures where all the
points are plotted with different symbols.} are:

\begin{displaymath}
\log{\SDM(\Re)} = 0.06 \log{\left ( L_{\rm V}/10^{11} \ {\rm
L_{\odot}} \right )} + 3.17 \ ,
\end{displaymath}

\begin{displaymath}
\log{\SDM(\Re)} = 0.08 \log{\left ( \mst/10^{11} \ {\rm M_{\odot}}
\right )} + 3.12 \ ,
\end{displaymath}

\begin{displaymath}
\log{\SDM(\Re)} = 0.05 \log \Re + 3.13 \ ,
\end{displaymath}

\begin{displaymath}
\log{\SDM(\Re)} = 0.10 \log{\left ( M_{200}/10^{12} \ {\rm
M_{\odot}} \right )} + 3.02 \ ,
\end{displaymath}
which are also shown in Fig.\,\ref{fig: bfseff} as compared to the
individual galaxy datapoints. The intrinsic scatter of the
fitting procedure, $\sigma_{\rm int}$, is formally negligible for all
the scaling relations above and will not to be considered further on,
while the true uncertainty of the fit
is dominated by the data point scatter, an issue we will come back later on.
As a virial mass estimate we use
$M_{200}$ (i.e., the mass within the radius $R_{200}$ where the
mean density is 200 times the cosmological mean matter density)
for homogenity with previous literature studies.

Taking at face values, these relations suggest a non universality
of $\SDM(\Re)$ in agreement with what we have found in CT10 using
a different halo model but the same reference radius. A similar
comparison with other results in literature is not straightforward
because of difference in the model adopted, the radius where
$\SDM(R)$ is evaluated for each individual system, and finally the
galaxy sample considered. E.g., using a NFW halo profile, Boyarsky
et al. (2009, hereafter B09) found a strong correlation between
$\SDM(R_s)$ and the halo mass over on wide range of masses,
ranging from dwarf spheroidal galaxies to galaxy clusters.
On the contrary, Donato et al. (2009, hereafter D09) using the
Burkert (1995, B95 hereafter) profile found a remarkably constant
$\SDM(R_c)$ ($R_c$ being the core radius of the B95 model) with
the luminosity on a sample of local spirals and ellipticals. These
``cored'' profile are in contrast, though, with our SIM model
which has no core but an inner cusp. Furthermore, as shown in
CT10, an inconvenience of the B95 model is that it does not allow
a decent fit to the SLACS data unless one assumes unreasonably low
values of the virial $M/L$ ratio.
Our lens systems, instead, is made of (mainly) ETGs at
intermediate $z$, thus partially overlapping with B09 and
completely complementary to the D09 sample. For these systems, we
cannot exclude that some of the discrepancy with the results above
can be advocated to some evolution with the redshift of the slope
of the $\SDM(\Re)$ vs $(L_V, \mst, \Re, M_{200})$. Albeit not
detailed investigated elsewhere (but see, e.g., B09 and NRT10 for
some hints), the galaxy morphology can be another important
responsible of the discrepancies in the scaling correlations
above. Our data are therefore not the best to solve the
differences between B09 and D09 contrasting results, however they
might provide a benchmark result for the correlations expected for
ETGs under a generalized halo model with no fixed cuspy density
profile (somehow covering the intermediate range between the B95
and the NFW central behavior).

This is particularly true due to the uncertainties on the slope
and zero-point of the fitted relations. Although we use a robust
Bayesian fitting method, the confidence regions around the
parameter fit are large (see Table \ref{tab: tabcorr}) due to the
large error bars on the DM related quantities. Such uncertainties
make any assessment on the non-universality of the relations
rather weak. Indeed, considering the $68\%$ confidence ranges, a
zero value for the slope can be statistically excluded only for
the $\SDM(\Re)$\,-\,$M_{200}$ relation, which is however
consistent with zero within $95\%$ CL.

Thus, looking at the correlations with the stellar quantities
(\Re, $M_*$, $L_V$), the $\SDM(\Re)$ universality cannot be ruled
out. On the other hand, the strong correlation with the halo mass
$M_{200}$ in Fig.\,\ref{fig: bfseff} works against the $\SDM(\Re)$
universality. This might be one reason for the contrasting results
between D09 and B09: in fact the former show the absence of a
trend of the column density with the luminosity, while the latter
show a strong trend with the halo mass, which is along the same
line of our conclusions.

However, even if we look at the correlations with luminosity, we
are inclined to argue that the claim of the constant $\SDM(\Re)$
with luminosity is motivated for the later type galaxies, while
the ETGs have an intrinsically larger scatter (see also NRT10)
which is the effect of a stronger correlation with the halo mass
(see also \S\ref{sec:gN}).

Focusing on the difference produced by the adoption of the
different halo profiles, we can now compare the results obtained
with the SIM model with findings in recent literature using the
same IMF and SLACS dataset.

CT10 have found a lower average DM column densities over the
sample, $\Delta \log{\SDM(\Re)} \sim 0.1 - 0.3$ with respect the
one obtained with the SIM. This is due to the fact that the SIM
model has an average central slope, $\gamma \sim 1.0 - 1.3$ which
is steeper than the one of the NFW and implies a projected DM mass
within \Re\ larger than the one obtained with the same NFW.

However, the slope of the $\SDM(\Re)$ vs $(L_V, \mst, M_{200})$
correlations stay very similar for the two halo models (see solid
black and red dashed lines in Fig.\,\ref{fig: bfseff})\footnote{We
note that in particular the correlation with $M_{200}$ might look
somehow too shallow to an eyeball check. We have checked that if
we use the sample F, the best fit slope increases to 0.13 which is
closer to the 0.17 value found when the NFW model is used (red
dashed) similarly to the correlation found with $L_V$ and \mst.
Moreover, a steeper slope could be found by excluding the least
massive points, although there seems to be nothing unusual for
these lenses to be reasonably excluded here and not elsewhere.}
which means that the details of the central DM density are
possibly insensible to the global quantities. On the contrary, the
$\SDM(\Re)$ seems much more sensitive to the scale of the luminous
matter as the slope of the $\SDM(\Re)$\,-\,$\Re$ correlation is
inverted with respect the NFW.
%
%
In particular, from Fig. 2, we see that the NFW produces almost
similar $\SDM(\Re)$ for smaller \Re\ (red dashed line) which means
that the two density profiles are very similar for more compact
objects, while it clearly produces much smaller DM column density
at larger \Re, where the SIM model accommodates the steeper
slopes. Despite the low significance of the trend, this is a
key result of our analysis because this points to a correlation
between the halo central slope and the size of the luminous matter
in the sense that bigger galaxies are formed in more cuspy halos.
This is qualitatively consistent with what already found in
DP09 and \cite{DPK09}, which have suggested that more massive
haloes are cuspier.  As a side note, the presence of steeper cusps
for massive systems goes in the same direction of the effect
expected for the canonical AC which seems to be necessary to model
extended kinematics, e.g., using planetary nebulae as a mass
tracers, in massive ETGs (\citealt{2011MNRAS.411.2035N}) and not
in more regular ones (\citealt{2009MNRAS.393..329N}). If so, the
steeper cusps would reduce the strength of the actual AC recipe
with respect the classical prescriptions (\citealt{Blumenthal+86};
\citealt{G04}) as reported elsewhere.

\begin{figure*}
\centering
\psfig{file=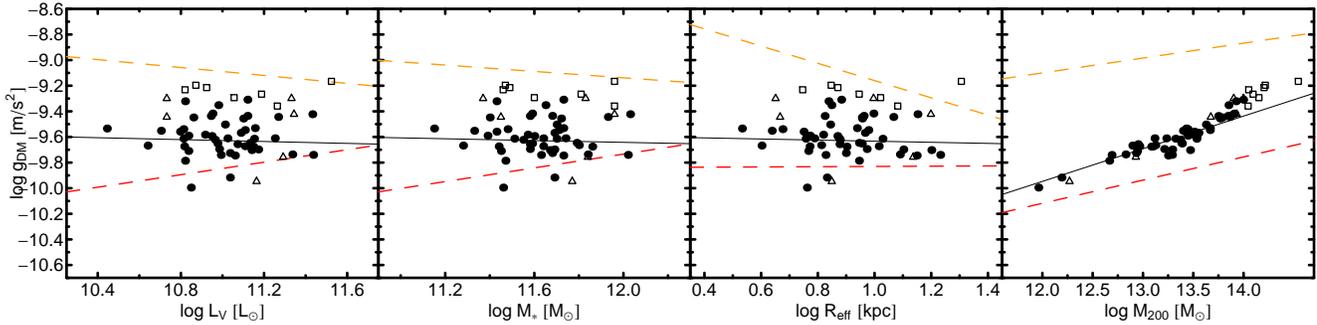, width=1\textwidth}\\
\caption{Best fit relations between the DM Newtonian acceleration
(in units of $m/s^2$) and the total luminosity $L_V$, stellar mass
\mst, effective radius \Re\ and halo mass $M_{200}$ from left to
right. Typical error bars are $\langle
\sigma[\log{\gDM(\Re)}]/\log{\gDM(\Re)} \rangle \simeq 2\%$.
Symbols and lines are the same as in Fig. \ref{fig: bfseff}.}
\label{fig: bfgdm}
\end{figure*}

Going to a comparison with more general mass profiles, T+10
adopted a singular isothermal model to describe the stellar\,+\,DM
mass profile. Here the difference is that the stellar masses have
been derived using different population models. The solid blue
lines in the two middle panels of Fig.\,\ref{fig: bfseff} shows
that their $\SDM(\Re)$\,-\,$\mst$ and $\SDM(\Re)$\,-\,$\Re$
relations have the opposite trend of the one we have found from
the SIM. Note that a decreasing $\SDM(\Re)$ with both the stellar
mass and the effective radius is also found by T+09 for local
ETGs, adopting the same mass model and stellar population
properties (gray continue line in Fig. \ref{fig: bfseff}), but
fitting very central velocity dispersions. A similar (albeit
shallower) decreasing trend and a lower average column densities
(of $\Delta \log{\SDM(\Re)} \sim 0.3 - 0.5$) are also found from
T+10 for a model with a constant M/L model (gray dashed line in
Fig. \ref{fig: bfseff}).

As T+10 do not assume a halo density profile, but derive the DM
mass by subtraction of the stellar component from the total
density profile, their approach seem much more dependent on the
stellar population analysis (i.e., IMF, priors on stellar
population parameters, availability of photometric or spectral
data, etc.). E.g., while stellar masses in T+10 are, on average,
consistent with the ones from A+09, some tilt is present and
finally propagates to a different final estimate of $\SDM(\Re)$.
Indeed, adopting the same total mass model, but the A+09 masses,
the $\SDM(\Re)$\,-\,$\mst$ correlations changes its sign, while
the $\SDM(\Re)$\,-\,\Re\ one becomes shallower so that the
disagreement with our SIM based results is partially
alleviated\footnote{Note that the results obtained in T+10, using
the two stellar mass sets, populate similar regions in the space
$\SDM(\Re)$\,-\,$\mst$ and $\SDM(\Re)$\,-\,$\Re$ (see cyan region
and the one enclosed within the blue lines in Fig. \ref{fig:
bfseff}).}. This check is a warning to us about the importance of
the mass model and stellar population analysis adopted which, in
principle, is an important player in the definition of the DM
scaling relations.

Keeping this in mind, we decided to check the scaling relations of
the total column density $\St$ (obtained including the stellar
mass within \Re) obtained with the SIM model for our particular
stellar population and IMF choice. As best fit correlations we
find\,:

\begin{displaymath}
\log{\St} = -0.17 \log{\left ( L_{\rm V}/10^{11} \ {\rm L_{\odot}}
\right )} + 3.44 \ ,
\end{displaymath}

\begin{displaymath}
\log{\St} = -0.16 \log{\left ( \mst/10^{11} \ {\rm M_{\odot}}
\right )} + 3.54 \ ,
\end{displaymath}

\begin{displaymath}
\log{\St} = -0.37 \log \Re + 3.74 \ ,
\end{displaymath}

\begin{displaymath}
\log{\St} = 0.01 \log{\left ( M_{200}/10^{12} \ {\rm M_{\odot}}
\right )} + 3.34 \ ,
\end{displaymath}
whose marginalized constraints on the slopes and the zeropoints
are given in Table \ref{tab: tabcorr}. A zero slope is excluded at
the $68\%$ CL for $\St$ vs $(L_V, \mst, \Re)$ correlations, while
we do not find any correlation with $M_{200}$. Thus, the net
effect of the inclusion of the stellar component is to tilt the
trend almost uniformly clockwise toward more negative slopes (and
contemporary increase the zeropoint due to the addition of the
stellar mass).
In fact, we have checked that the slope in the central regions of
our light profiles become shallower at larger \Re, possibly
combining with the DM slope the results above. One of the possible
driver of such trends is galaxy merging, which has been shown to
produce shallower slopes in simulated haloes (e.g.,
\citealt{BM04}).
The trend of the \St\ seems consistent with the recent finding
from \cite{Auger+10} which, adopting a simple power-law density
to shape the full mass profile of the SLACS lenses,
have found a strong inverse correlation of the total density
slope with \Re.

\subsection{Newtonian acceleration}\label{sec:gN}

The issue of the universality of the column density $\SDM(R)$ has
some important dynamical consequences if one considers that it can
be easily related to the Newtonian acceleration $\gDM(r) = G
M_{\rm DM}(r)/r^2$ (Gentile et al. 2009, hereafter G09). In
particular, G09 have shown evidences for the universality of both
$\gDM(R_c)$ and $g_{\star}(R_c)$ with the label $DM$ ($\star$)
referring to DM (stellar) quantities, over a (small) sample of
spirals and ellipticals. Let us first discuss the DM case
evaluating the acceleration at the effective radius to be
consistent with our choice throughout the paper. For the best fit
relations (plotted in Fig.\,\ref{fig: bfgdm}), we get\,:

\begin{displaymath}
\log{\gDM(\Re)} = -0.04 \log{\left ( L_{\rm V}/10^{11} \ {\rm
L_{\odot}} \right )} - 9.63 \ ,
\end{displaymath}

\begin{displaymath}
\log{\gDM(\Re)} = -0.03 \log{\left ( \mst/10^{11} \ {\rm
M_{\odot}} \right )} - 9.61 \ ,
\end{displaymath}

\begin{displaymath}
\log{\gDM(\Re)} = -0.04 \log \Re - 9.60 \ ,
\end{displaymath}

\begin{displaymath}
\log{\gDM(\Re)} = 0.25 \log{\left ( M_{200}/10^{12} \ {\rm
M_{\odot}} \right )} - 9.95 \ ,
\end{displaymath}
where the accelerations are in units of $m/s^2$. Similarly to the
column densities, we find that the DM Newtonian acceleration has
no correlation with the stellar quantities $(L_V, \mst, \Re)$,
while it strongly correlates with $M_{200}$. Thus we confirm that
\gDM\ is a constant with respect to the stellar quantities
although it is larger than the value found by G09 for the \gDM\
computed at $R_c$. The latter is an upper limit to our \Re\
estimates since generally $R_c \gsim \Re$ and for a given constant
core density $\rho_0$ the Newtonian acceleration scales linearly
with radius, i.e. $\gDM(\Re)\lsim \gDM(R_{\rm c})$. As already noted
by NRT10, this result works against the universality of the \gDM,
which instead seems to scale with the morphological type (and
possibly the mass).

The SIM estimates obtained here are discrepant with the results
from CT10 for the NFW and B95 (Fig.\,\ref{fig: bfgdm}). In
particular our \gDM\ are located in between the two reference halo
models, while the difference in the slopes with all the quantities
are statistically insignificant. This result is somehow different
with respect the $\SDM(\Re)$ in Fig. 2 where the CT10 estimates
are almost everywhere lower than the SIM estimates, which might be
tracked to the fact that here we are considered 3D quantities. In
particular, the large values obtained by CT10 for the B95 are
related to the fact that typical core radius in CT10 are smaller
than \Re\ thus not representing a real cored profile, but rather
more resembling a pseudo-isothermal sphere, with a rather steep
slope around \Re. In fact the deviations are more marked for the
smaller systems which shall have also shallower inner slopes
(according to Eq. 5), and thus are overestimated by the steeper
densities implied by the B95 models. On the other hand, the
$\gDM(\Re)$ are systematically larger than the values obtained in
CT10 for NFW because of the steeper 3D slopes. We remark here that
the 3D quantities are more weakly depending on the inner slopes
with respect the projected ones shown in Fig. 1, a fact that
translates with an null correlation of the \gDM\ with \Re.

Going to the stellar Newtonian acceleration, we find as best fit relations\,:

\begin{displaymath}
\log{g_{\star}(\Re)} = -0.42 \log{\left ( L_{\rm V}/10^{11} \ {\rm
L_{\odot}} \right )} - 9.39 \ ,
\end{displaymath}

\begin{displaymath}
\log{g_{\star}(\Re)} = -0.35 \log{\left ( \mst/10^{11} \ {\rm
M_{\odot}} \right )} - 9.18 \ ,
\end{displaymath}

\begin{displaymath}
\log{g_{\star}(\Re)} = -1.03 \log \Re - 8.48 \ ,
\end{displaymath}

\begin{displaymath}
\log{g_{\star}(\Re)} = -0.07 \log{\left ( M_{200}/10^{12} \ {\rm
M_{\odot}} \right )} - 9.31 \ ,
\end{displaymath}
which clearly demonstrate that this is not a universal quantity
(and indeed the confidence ranges for the slope in Table \ref{tab:
tabcorr} exclude a zero slope at the $68\%$ CL) in agreement with
CT10. Note, however, that a correlation of $g_{\star}(\Re)$ with
the stellar quantities $(L_V, \mst, \Re)$ is expected, being all
these quantities involved in its definition. On the contrary, the
correlation with $M_{200}$ is very weak, which shows that there is
not any strong dependence of the stellar mass in the centers on
the global DM content.
%

\subsection{Dark matter mass content}

In the previous sections we have seen that the total and DM column
densities correlate differently with luminosity, stellar mass and
\Re\ which shall depend on the different spatial distributions of
stellar and dark matter. Here we want to investigate the
consequences of the column density scaling relations in term of
the central DM fractions. To this end, we note that

\begin{displaymath}
\frac{\SDM}{\St} = \frac{M_{\rm DM}^{\rm proj}(\Re)}{\mst^{\rm
proj}(\Re) + M_{\rm DM}^{\rm proj}(\Re)} = f_{DM}^{\rm proj}(\Re)
\ ,
\end{displaymath}
where $f_{\rm DM}^{\rm proj}(\Re)$ is the projected DM fraction.
Considering the best fit values of the correlations of both column
densities with stellar parameters, we obtain that the projected DM
fraction within \Re\ scales with both the luminosity, stellar mass
and \Re, as $f_{\rm DM}^{\rm proj}(\Re) \propto L_V^{0.23}$,
$f_{\rm DM}^{\rm proj}(\Re) \propto \mst^{0.24}$ and $f_{\rm DM}^{\rm
proj}(\Re) \propto \Re^{0.42}$.

The same results are found by direct fitting of the same
quantities of the individual galaxies, adopting our Bayesian
fitting procedure\,:

\begin{displaymath}
\log{f_{\rm DM}^{\rm proj}(\Re)} = 0.21 \log{\left ( L_V/10^{11} \ {\rm
L_{\odot}} \right )} - 0.21 \ ,
\end{displaymath}

\begin{displaymath}
\log{f_{\rm DM}^{\rm proj}(\Re)} = 0.18 \log{\left ( \mst/10^{11} \ {\rm
M_{\odot}} \right )} - 0.32 \ ,
\end{displaymath}

\begin{displaymath}
\log{f_{\rm DM}^{\rm proj}(\Re)} = 0.41 \log{\Re} - 0.58 \ ,
\end{displaymath}
while the $68\%$ confidence ranges of the slopes are $(0.15,
0.26)$, $(0.12, 0.24)$ and $(0.34, 0.45)$ for the
$f_{\rm DM}^{\rm proj}$\,-\,$L_V$, $f_{\rm DM}^{proj}$\,-\,$\mst$ and
$f_{\rm DM}^{\rm proj}$\,-\,\Re, respectively. We have therefore a
clear evidence that the different scalings of $\SDM(\Re)$ and
$\St$ with the stellar quantities are an expected consequence of
the varying DM content within the effective radius.

Similarly, we define the three dimensional DM fraction as
$f_{\rm DM}(\Re) = M_{\rm DM}(\Re)/[\mst(\Re) + M_{\rm DM}(\Re)]$ and
investigate its correlation with the stellar quantities. Our best
fit relations are\,:

\begin{displaymath}
\log{f_{\rm DM}(\Re)} = 0.22 \log{\left ( L_{\rm V}/10^{11} \ {\rm
L_{\odot}} \right )} - 0.44 \ ,
\end{displaymath}

\begin{displaymath}
\log{f_{\rm DM}(\Re)} = 0.15 \log{\left ( \mst/10^{11} \ {\rm
M_{\odot}} \right )} - 0.53 \ ,
\end{displaymath}

\begin{displaymath}
\log{f_{\rm DM}(\Re)} = 0.59 \log \Re - 0.98 \ ,
\end{displaymath}
while the $68\%$ CL for the slope are $(0.08, 0.30)$, $(0.02,
0.24)$, $(0.36, 0.71)$ respectively. Despite the large
uncertainties, these results show that brighter, more massive and
bigger systems have a larger DM content within the effective
radius in qualitative agreement with previous results in
literature, regardless the adoption of deprojected or projected
quantities (e.g. \citealt{Pad04}; \citealt{Cappellari+06}; T+09;
\citealt{Auger+10}; NRT10; T+10). We, however, note that our best
fits relation with $L_V$ and $\mst$ are much shallower than what
is found in CT10 for the fiducial NFW model (and Salpeter IMF),
but in good agreement with the ones in \cite{C09}, where a
phenomenologically motivated general halo profile was used to fit
the same SLACS data considered here.

\subsection{Impact of the IMF choice}

All the results discussed so far has been obtained using the SIM model for the dark halo and a Salpeter IMF. As discussed in Sect.\,3.3, choosing a Chabrier IMF leads to SIM models having a virial mass larger than $10^{14} \ {\rm M_{\odot}}$ for more than $50\%$ of the sample so that we have preferred to exclude the SIM\,+\,Chabrier combination when discussing the scaling relations.
This is a reasonable choice under the hypothesis of a universal IMF which
does not depend on galaxy parameters. There are different evidences that
this might not be the case, although there is still not any consensus whether
the IMF might change with galaxy morphology, luminosity/mass and/or stellar population parameters \citep{D08,vD08,H10,NRT10,vDC10,vDC11}.
Here we decided to check what can be the
impact of the IMF choice, by computing the scaling relations for a mixed sample
made out of the 25 lenses in the G sample of the SIM\,+\,Chabrier model and the
remaining 26 lenses of the G sample of the SIM\,+\,Salpeter model. In a sense,
we are here postulating a IMF varying with
the halo virial mass and approximating such a variation with a rough step function.

Let us consider first the DM column density within $R_{\rm e}$. For the best fit relations, we get\,:

\begin{displaymath}
\log{{\cal{S}}_{\rm DM}(R_e)} = -0.003 \log{(L_V/10^{11} \ {\rm L_{\odot}})} + 3.21 \ ,
\end{displaymath}

\begin{displaymath}
\log{{\cal{S}}_{\rm DM}(R_e)} = -0.02 \log{(M_{\star}/10^{11} \ {\rm M_{\odot}})} + 3.22 \ ,
\end{displaymath}

\begin{displaymath}
\log{{\cal{S}}_{\rm DM}(R_e)} = -0.09 \log{R_e} + 2.33 \ ,
\end{displaymath}

\begin{displaymath}
\log{{\cal{S}}_{\rm DM}(R_e)} = 0.06 \log{(M_{200}/10^{12} \ {\rm M_{\odot}})} + 3.11 \ .
\end{displaymath}
Comparing with the values in Table 1, we see that the best fit relations are shallower for $(L_V, M_{\star})$ and steeper for $(R_e, M_{200})$. However, if we consider the $68\%$ confidence ranges, the slopes are fully consistent so that the change in the slope can not be considered statistically significant.

Similarly, the total column density ${\cal{S}}_{\rm tot}$ turned out to have the
best fit relations with slopes $(-0.21, 0.05, -0.04, -0.04)$ for $(L_V, M_{\star}, R_e, M_{200})$ respectively. Also for these relations the confidence ranges significantly overlap with respect to the Salpeter IMF (see Table 1) thus
the two cases do not differ significantly.

For the DM Newtonian acceleration we obtain\,:

\begin{displaymath}
\log{g_{\rm DM}(R_e)} = -0.05 \log{(L_V/10^{11} \ {\rm L_{\odot}})} - 9.45 \ ,
\end{displaymath}

\begin{displaymath}
\log{g_{\rm DM}(R_e)} = -0.07 \log{(M_{\star}/10^{11} \ {\rm M_{\odot}})} - 9.42 \ ,
\end{displaymath}

\begin{displaymath}
\log{g_{\rm DM}(R_e)} = -0.13 \log{R_e} - 9.33 \ ,
\end{displaymath}

\begin{displaymath}
\log{g_{\rm DM}(R_e)} = 0.13 \log{(M_{200}/10^{12} \ {\rm M_{\odot}})} - 9.67 \
\end{displaymath}
which are comparable with the results for slope and zeropoint as in Table 1.
This suggests that the universality (or lack of) of the DM Newtonian acceleration
is not significantly affected by the choice of the IMF (a similar results is found for the stellar Newtonian acceleration which we do not report for brevity).

Finally, we have considered the DM mass fraction within $R_e$ and found:

\begin{displaymath}
\log{f_{\rm DM}(R_e)} = 0.18 \log{(L_V/10^{11} \ {\rm L_{\odot}})} - 0.30 \ ,
\end{displaymath}

\begin{displaymath}
\log{f_{\rm DM}(R_e)} = -0.10 \log{(M_{\star}/10^{11} \ {\rm M_{\odot}})} - 0.24 \ ,
\end{displaymath}

\begin{displaymath}
\log{f_{\rm DM}(R_e)} = 0.44 \log{R_e} - 0.70 \ .
\end{displaymath}

As expected, these results turned out to be significantly different from the
Salpeter IMF case.
In particular, the $f_{\rm DM}$\,-\,$L_V$ and $f_{\rm DM}$\,-\,$R_e$ relations turned
out to be shallower, while the $f_{\rm DM}$\,-\,$M_{\star}$ has a negative slope and
a significant anti--correlation. This is the consequence of the step IMF function
assumption that produced a significant increase of the $f_{\rm DM}(R_e)$ for the
lower mass systems with the Chabrier with respect the smaller $f_{\rm DM}(R_e)$
obtained with the Salpeter IMF assumed for the more massive ones. The same
argument applies to the $f_{\rm DM}$\,-\,$L_V$ and $f_{\rm DM}$\,-\,$R_e$ relations
which turned out to be shallower.

As a final remark, we can conclude that scaling relations are overall mildly
affected by the IMF assumption in the dynamical/lensing analysis, with
correlations being comparable among results obtained either assuming a
universal (Salpeter) IMF, or a non universal mass dependent IMF (Chabrier for
less massive systems and Salpeter for more massive ones).
A universal Chabrier IMF seems to be ruled out because it produces too many
systems with unrealistically large virial masses for galaxy systems.

\section{Conclusions}

While there is a general consensus on the ubiquitous presence of
DM in galaxies, there is still an open debate about its mass
density distribution is and its role it has exactly played in the
galaxy formation scenario. These are crucial issues since the
exact density distribution and the assembly processes that dark
halos have undergone through might tell more on the actual nature
of the DM itself. Scaling relations among DM related quantities
and stellar properties may give important hints about the relative
interplay between the two main constituents of galaxies.

In this context, it seems particular intriguing the existence of some
universal properties of the DM quantities like the column density
or even a presence of a characteristic acceleration scale (e.g.
\citealt{D09,G09}) which might be related to common formation
processes on many mass scales or morphological categories. Recent
works have argued either in favor or against the presence of such
universal values (B09, CT10, NRT10). Dealing with DM properties,
though, means necessarily to deal with indirect, model dependent
quantities, which can make the conclusions on these parameters
strongly affected by the particular model adopted.
One approach might be to use models
which bracket the widest range of DM properties derived from the
N-body cosmological simulations, e.g. from the cuspy profiles
predicted by the classical NFW to the ``cored'' profiles of the
B95 (see e.g. CT10). The disadvantage of this approach is that it
gives a partial vision of how the DM scaling relations vary in the
two extreme regime without allowing to generalize the results in
case the actual DM properties are in the between.

In the attempt of testing more general, theoretical motivated DM
halo profiles, like the one proposed by the secondary infall
model as implemented in DP09, we have modeled the central velocity
dispersion and the projected mass within the Einstein radius of a
large sample of ETGs lenses at intermediate redshift $(\langle z
\rangle \sim 0.2)$.

In particular, the DP09 models adopted a modified secondary
infall scenario including (in a semianalytic way) the effect of
angular momentum, dynamical friction and adiabatic collapse of
baryons. As a first important result, we have shown that this
model is fully compatible with observations of ETG and well performing
in the data fitting.
%
%
This is a significant step forward with respect to previous analyses
where SLACS sample have been modeled with standard NFW (CT10),
as the SIM model has the main property to be fully assigned by a
single parameter (the virial mass) instead of the two parameters
required by more ``standard'' halo density models like NFW
(which is specified once concentration and virial mass are given)
or Burkert profile (characterized by the core radius and the
central density).

Even if we consider the correlations between the halo parameters for
the NFW \citep{NFW96,B01,H07} and Burkert profiles
(e.g. Salucci \& Burkert 2000) which allow us to re-write these models
as a function of one parameter only (e.g. the virial mass), we have seen
that the best--fit to the observed quantities turned out to be generally
poorer than the one provided by the SIM model (see Sect. \ref{sec:SIMan}),
with only NFW providing somehow similar significance than the SIM model.
This is mainly because in the typical mass range spanned by the ETG sample
considered here, the SIM model predicts cuspier profiles according with
Eq.(\ref{eq: gammamv}), i.e. NFW--like or even cuspier.

With this novel DM setup, we have estimated the DM column density
$\SDM(\Re)$, the Newtonian acceleration $\gDM(\Re)$ and, finally
the DM mass fraction $f_{\rm DM}(\Re)$ along with their correlations
with the stellar total luminosity, mass and size. The best fit
relations show that $\SDM(\Re)$ is almost constant (possibly
increasing with, if any) over the range of stellar parameters
$(L_V, \mst, \Re)$ probed by the adopted dataset, while it is
clearly strongly correlated with the halo mass $M_{200}$. This
result is actually consistent with both G09, claiming a
characteristic density scale over a large range of galaxy
luminosities, and with B09 which instead has found a correlation
with the dark halo mass.

Similarly we have found an even more remarkably constant DM
Newtonian acceleration $\gDM(\Re)$  with $(L_V, \mst, \Re)$, and
still a strong correlation of this quantity with the virial mass.
In this case, though, the absolute constancy of the $\gDM(\Re)$
does not allow to justify the correlation with the virial mass
%
%
but probably says more of some intrinsic properties of the dark
matter halos, and of the non--existence of an universal
acceleration scale. In fact, we have confirmed here a former
evidence from NRT10, that $\gDM(\Re)$ of the ETGs is on average
larger than the one obtained for late-type galaxies (as e.g. in
Donato et al. 2009, having considered that their column density
obtained at the core radius are an upper limit for the same
quantity if computed at the \Re).

As such, one could argue of the existence of a Newtonian
acceleration growing with the morphological type and (possibly)
with the stellar mass (if including all the Hubble sequence) as
well as shown by the trend with $M_{200}$.

We need to conclude this reasoning with a caveat, underlying all
analyses based on model dependent approaches: all DM quantities
correlations against $(L_V, \mst, \Re)$ are critically dependent on
the adopted halo model and stellar IMF (in our case a Salpeter
IMF). A direct comparison with previous literature results is
complicated by systematic effects due to differences in both the
halo model (SIM vs NFW or Burkert) and the radius where the DM
quantities are evaluated ($\Re$ vs $R_s$ or $R_c$). In particular,
the comparison of results obtained with the SIM model with similar
works based on the adoption of the more standard NFW and B95 has
highlighted an interesting correlation between the cuspyness of
the dark halo and the size of the parent galaxies. In the SIM
approach, in fact, larger galaxies seem to assemble in dark halo
having steeper cusps, while more compact massive ETGs are
accommodated on shallower cusps. This is an interesting hint that
might be cross-checked on hydrodynamical simulations and seems to
be a crucial test for the hierarchical model as a whole.

Finally we stress that there are still some weaknesses in the
analysis proposed, mainly posed by the limited galaxy sample (e.g.
the SLACS sample only probes a quite limited range in both stellar
luminosity and mass), and the large uncertainties on the derived
quantities. However we have considered this a benchmark test for
more general dark halo model laws based on a simple physically
motivated galaxy model. This is the first step of a broader plan
to improve the testing along different roads. On one hand, a
detailed theoretical investigation is needed to find out
quantities which depends as less as possible on both the adopted
halo profile and stellar IMF. Similarly, one should check whether
the choice of \Re\ as a reference radius where DM quantities are
evaluated is the most convenient one finding a compromise between
the need to not extrapolate outside regions directly probed by
data and the halo model characteristics. On the other hand,
stronger constraints on the slope of the investigated correlations
could be obtained by narrowing the uncertainties on the DM
quantities. The use of a one parameter model, like the SIM, is
expected to help by eliminating the degeneracies among halo
parameters which generally plague the analysis adopting the NFW
profiles or similar, and contribute to the overall error budget. A
further improvement would be obtained fitting the full velocity
dispersion profile rather than its aperture value only, although
such a strategy could be applied to local ETGs only. Finally, a
larger sample spanning a wider range in $(L_V, \mst, \Re,
M_{200})$ would allow to further narrow down the confidence ranges
for the slopes of the investigated correlations by both improving
statistics and better tracking the different trends.

Should both constancy with stellar luminosity, mass and size and
independence on the fitting procedure details be successfully
demonstrated, one could safely conclude that a proposed quantity
is indeed universal and use such a result to constrain galaxy
formation and evolution scenarios.

\section*{Acknowledgments}

VFC and CT are funded by the Italian Space Agency (ASI) and the Swiss National Science Foundation, respectively.

\end{document}